\documentclass{article}
\usepackage{amssymb,amsmath}
\usepackage[margin=1.5in]{geometry}
\usepackage[utf8]{inputenc}
\usepackage{graphicx} 
\usepackage[sorting=none]{biblatex}

\usepackage{xcolor}
\usepackage{braket}
\usepackage{subcaption}
\usepackage{multirow}

\usepackage{hyperref}

\newcommand{\defeq}{\overset{\text{\tiny def}}{=}}

\title{Alternative quantisation condition for wavepacket dynamics in a hyperbolic double well}
\author{D. Kufel, H. Chomet, C. Figueira de Morisson Faria\\}
\date{
Department of Physics and Astronomy, University College London,\\ Gower Street, London WC1E 6BT, UK \\[2ex]
\today
}

\begin{document}
\maketitle

\begin{abstract}
We propose an analytical approach for computing the eigenspectrum and corresponding eigenstates of a hyperbolic double well potential of arbitrary height or width, which goes beyond the usual techniques applied to quasi-exactly solvable models. We map the time-independent Schr\"odinger equation onto the Heun confluent differential equation, which is solved by using an infinite power series. 
The coefficients of this series are polynomials in the quantisation parameter, whose roots correspond to the system's eigenenergies. 
This leads to a quantisation condition that allows us to determine a whole spectrum, instead of individual eigenenergies. This method is then  employed to perform an in depth analysis of electronic wave-packet dynamics, with emphasis on intra-well tunneling and the interference-induced quantum bridges reported in a previous publication [H. Chomet et al, New J. Phys. \textbf{21}, 123004 (2019)]. Considering initial wave packets of different widths and peak locations, we compute autocorrelation functions and Wigner quasiprobability distributions. Our results exhibit an excellent agreement with numerical computations, and allow us to disentangle the different eigenfrequencies that govern the phase-space dynamics.
\end{abstract}

\section{Introduction}
\label{sec:intro}

Analytical modeling is widely used in many areas of physics. It provides key insight and interpretational power, which may be unavailable in purely numerical approaches. Although numerical models are versatile and extremely useful for quantitative comparisons, the physics involved may be difficult to extract. For that reason, analytic models are employed to establish paradigms, or distill the essential features of a physical system. 
For instance, the harmonic oscillator is widely used in many areas of physics, such as quantum optics, solid state physics, or molecular physics to describe modes of the electromagnetic field, lattice or molecular vibrations (see for example \cite{QHOreview}). Within strong-field laser-matter interaction, the Gordon-Volkov solution \cite{Gordon1926,Volkov1935} has been widely used approximate the electron dynamics by field-dressed plane waves. This solution is exact and constitutes an important ingredient in constructing the strong-field approximation, which is a  semi-analytic approach that can be linked to interfering electron orbits \cite{Amini2019}. An orbit-based interpretation was vital to the description of strong-field phenomena as the laser-induced rescattering or recombination of an electron with its parent ion \cite{Corkum1993,Lewenstein1994,Becker2018}. This interpretation led to the inception of attosecond science \cite{Lein2007,Krausz2009,Salieres2012R}, which may allow steering electron dynamics in real time. Attoscience is a particularly challenging area as the Hamiltonians are time dependent and the phenomena highly transient. Thus, analytical solutions are either inexistent or hard to find. Furthermore, perturbation theory with regard to the field is not applicable.  

In contrast, if the Hamiltonian does not vary with time, analytically solving the time-dependent Schr\"odinger equation (TDSE) reduces to an eigenvalue problem. 
The challenge is then to solve the time-independent Schr\"odinger equation (TISE) for a potential of interest and to find the time-independent eigenfunctions. In principle, describing the temporal evolution of a wavepacket in an eigenstate basis is very convenient, as it boils down to computing the (time-independent) overlap integrals of the initial wave packet with the bound states and inputting the phase factors  $\exp(-i E_n t / \hbar)$, where $E_n$ denotes the system's  eigenenergies. In practice, however, computing overlap integrals may not be an easy task, and may even not be feasible. In fact, the number of exactly solvable problems in quantum physics is quite limited, even for simplified cases. Many widespread models, such as the one-dimensional soft-core potentials, have no analytical solution. 

The dearth of analytically solvable models in physics has led to the development of quasi-exactly solvable (QES) analytical models \cite{ushveridze,turbiner}.
The QES exploit the fact that Hamiltonian operator (from a quasi-exact subclass) may be represented as an infinite dimensional \textit{block diagonal} matrix\footnote{This is in contrast to exactly-solvable models for which infinite-dimensional matrix representation of the Hamiltonian is \textit{diagonal} and the \textit{entire} spectrum of eigenvalues may be found.}, hence allowing to explicitly find a certain \textit{subset} of all eigenfunctions and eigenvalues by diagonalising one of the finite-dimensional blocks \cite{ushveridze}. Unfortunately, QES may be of limited practical use. In particular, to obtain such subset of eigenvalues, it is necessary to constrain the parameters of the potential, such that certain infinite power series terminates to a polynomial \cite{downing2013,}. Such parameter-constrained potential may not be physically relevant. Case in point, the constrained potential may effectively represent two almost detached wells, where most of the interesting physics is not captured. Furthermore, it is not guaranteed that the found eigenvalue subset will contain the energy range of interest. Finally, it may be that the physical problem requires the knowledge of the whole eigenspectrum. For instance, a small eigenvalue subset may not be advantageous for calculating a large enough number of overlap integrals in order to accurately determine the time evolution of the wavepacket.

Here we propose a one-dimensional analytical method to investigate the dynamics of a wavepacket in a field-free, hyperbolic double-well potential in one dimension - using the potential form proposed by \cite{downing2013} as a case study. Double well potentials serve as toy models for molecular systems (such as electronic wavefunction in $H_2^{+}$ \cite{harrell1980double} or nitrogen inversion in $NH_3$ \cite{davis1973analytical}, semi-conductor heterostructures \cite{alferov2001nobel} and optical lattices \cite{schumm2005matter}. 

Rather than exactly truncating the infinite power series to a polynomial - thus making the model to be quasi-exactly solvable - we instead propose an alternative quantisation condition for such potential. This condition reduces the problem of obtaining admissible energies to finding the roots of a polynomial generated from a three-term recurrence relation. By these means we obtain the entire spectrum of eigenenergies - which may be found to an arbitrary level of precision - for any values of depth and peak locations of the potential. We analytically evaluate the overlap integrals for the appropriately designed wavepackets as functions of their parameters, thus allowing to predict their time evolution in various setups. It appears that the methods developed here may be applicable also to a much wider class of the hyperbolic potentials for which Schr\"odinger's equation reduces to Heun's equation - in particular we show that the quantisation condition proposed here correctly predicts the eigenvalues for the potentials proposed earlier by \cite{manning,xie2012new,hartmann2014}.

The approach developed here is then applied to molecular tunneling. This is motivated by recent studies of strong-field enhanced ionisation in stretched molecules, in which momentum gates in phase space have been identified using Wigner quasiprobability distributions. These gates allow a direct intra-molecular population flow and were attributed to the system's non-adiabatic response to a a strong laser field \cite{takemoto2011}. Recently, however, we have shown that momentum gates also exist for static fields, or even in the field-free case \cite{chomet2019}. The key physical mechanism facilitating such gates is quantum interference, which provides a bridge for the electronic wavepacket to reach the other centre and ultimately the continuum. These quantum bridges perform a clockwise rotation in phase space, whose frequency depends on the initial wavepacket and the internuclear separation. However, it is yet not understood what properties of the system determine these frequencies. The analytical model developed in this work is ideally placed for an in-depth study of how the initial electronic wavepacket influences this motion, and how it is related to the system's eigenfrequencies. Specifically, a hyperbolic double well potential has several desirable properties for the molecular toy model. First, the limit $V(x) \rightarrow 0$ as $x \rightarrow \pm \infty$ allows for the existence of continuum of states for positive electron energies.  This is in contrast to the models of double-well potential by e.g. \cite{razavy1980exactly} or \cite{konwent1995certain} for which $V(x) \rightarrow \infty$ as $x \rightarrow \pm \infty$. Second, it 
allows to faithfully model the binding potential in the region of interest (i.e. close to the central barrier \cite{chomet2019}). Third, the location of the (symmetric) wells and peak value of $V(x)$ may be independently tuned. Finally, although other hyperbolic double-well models such as those developed by \cite{xie2012new} or \cite{hartmann2014} can reliably model the central potential barrier, the one we are using leads to an impenetrable barrier by classical means. This is important to rule out other population-transfer mechanisms. 

This article is organised as follows. In Sec.~\ref{sec:symmetric_treatment}, the method developed by us is outlined, including how the Schr\"odinger equation can be reduced to Heun's equation (Sec.~\ref{sub:reductionofschrodingerstoheuns}), the quantisation condition we propose (Sec.~\ref{sub:quantisationcondition}), determining the number of bound states (Sec.~\ref{sub:numbBoundStates}) and how to construct appropriate wave packets and ascertain their time evolution (Sec.~\ref{sub:wavepackets}). Subsequently, in Sec.~\ref{sec:applications}, we apply the model to tunneling dynamics, analysing the time profiles of autocorrelation functions (Sec.~\ref{sec:autocorr}) and Wigner quasiprobability distributions (Sec.~\ref{sec:Wigner}). In particular, we determine the main frequencies with which the above-mentioned quantum bridges propagate and their dependence on the initial wave packet. Finally, in Sec.~\ref{sec:conclusion}, we conclude the paper and discuss possible future directions. 

\section{Methods \label{sec:symmetric_treatment}}

We will consider the evolution of a time-dependent wave packet $\Psi(x,t)$ using the basis of eigenstates $\psi_n(x)$ that solve the time-independent Schr\"odinger equation (TISE)
\begin{equation}
    \hat{H} \psi_n(x)=E_n \psi_n (x), 
\end{equation}
with the Hamiltonian defined by
\begin{equation}
    \hat{H}=\frac{\hat{p}^2}{2m}+ V(x). 
\end{equation}
The binding potential
\begin{equation}
V(x)=-V_0 \frac{\sinh^{4}(x/d)}{\cosh^{6}(x/d)}
\label{potential_generic}
\end{equation}
is a member of the wider family of symmetric hyperbolic potentials of the form
\begin{equation}
    V^{(m)}(x)=-V_0^{(m)}\frac{\sinh^{2m}(x/d)}{\cosh^{2m+2}(x/d)},
\end{equation}
where $V_0^{(m)}$ specifies the depth of the potential and $d$ its peak location. For $m=1,2$ they produce a double-well (bistable) potential and for $m=0$ they reduce to the (single-well) P\"oschl-Teller potential, which is exactly solvable \cite{poschlteller1933}. 

In the eigenbasis of the TISE, 
\begin{equation}
    \Psi(x,t)= \sum_n \Lambda_n \exp(-i E_n t / \hbar)\psi_n(x),
    \label{eq:wpcoherent}
\end{equation}
 where
 \begin{equation}
     \Lambda_n=\int \Psi(x,0) \psi^*_n(x) dx
     \label{eq:overlapintegrals}
 \end{equation}
 are overlap integrals between the initial wavepacket $\Psi(x,0)$ and eigenfunctions $\psi_n(x)$ of the hyperbolic double well. The goal of the present investigation is to determine $\Lambda_n$ by analytical means.
 
The 1D time-independent Schr\"odinger equation (TISE) for the potential given in Eq.~(\ref{potential_generic}) reads as
\begin{equation}
    \frac{d^2\psi(z)}{dz^2}+\left(\epsilon d^2 +U_0 d^2 \frac{\sinh^4(z)}{\cosh^6(z)}\right)\psi(z)=0,
\label{schrodinger_generic}
\end{equation}
with dimensionless parameters $z=x/d$, $U_0=2mV_0/\hbar^2$, $\epsilon=2mE/\hbar^2$. Note that the potential $V(x)$ has even parity, which implies the existence of even/odd parity wavefunctions. 

\subsection{Reduction of Schr\"odinger's to Heun's equation \label{sub:reductionofschrodingerstoheuns}}

It was shown \cite{downing2013} that for even parity wavefunctions the above equation may be reduced to the Heun confluent differential equation by introducing the new variable $\xi=1/\cosh^2 z$ (with $0<\xi \leq 1$ as $-\infty<z<+\infty$), that is, 
\begin{equation}
\frac{d^2}{d\xi^2} H(\xi) + \left(\alpha + \frac{\beta + 1}{\xi}+\frac{\gamma+1}{\xi-1}\right)\frac{d}{d\xi}H(\xi)+\left( \frac{\mu}{\xi} +\frac{v}{\xi-1} \right) H(\xi) = 0,
\label{Heun_generic}
\end{equation}
where the even-parity solutions to TISE are of the following form:
\begin{equation}
  \psi_{even}(\xi)=\xi^{\beta/2} e^{\alpha \xi/2} H(\alpha,\beta,\gamma,\delta,\eta,\xi)  ,
\label{wavefunctions_even}
\end{equation}
with $\alpha=-d\sqrt{U_0}<0$, $\beta=-id\sqrt{\epsilon}>0$ and others given by $\gamma=-\frac{1}{2}$, $v= \frac{1}{4} \left( \alpha+\beta(\beta+1) \right)$, $\delta=\mu+v-\frac{\alpha}{2} \left( \beta + \gamma +2 \right)$, $\mu=\frac{1}{4} \left( \alpha (\alpha+2)+2\alpha \beta - \beta (\beta+1) \right)$, $\eta=\frac{\alpha}{2} \left( \beta +1 \right) - \mu - \frac{1}{2} \left( \beta + \gamma + \beta \gamma \right)$. 

At this point we should make few technical comments. First, note that the $x \rightarrow \xi$ mapping is not injective and uniquely represents only the half of the $-\infty < x < +\infty$ \ range. However, the other half of the range is just the symmetric copy of the first one - this mapping intrinsically constrains the wavefunctions to be even in $x$-variable space. Second, as Heun's confluent equation arises from Schrodinger's equation, some of the parameters out of $\alpha$, $\beta$, $\gamma$, $\delta$, $\mu$, $v$ are dependent on each other: in fact only $\alpha$ (property of the depth and width of the potential) and $\beta$ (free parameter which determines the allowed energies) are independent.

Hence, we may write $H(\alpha, \beta, \gamma, \delta, \eta, \xi)$ as the following infinite power series involving only two parameters\footnote{From the practical point of view, different notations for parameters of confluent Heun's function are used. The convention used in this paper may be converted to the one in HeunC[$q'$,$\alpha'$,$\gamma'$,$\delta'$,$\epsilon'$] function in Mathematica using $\epsilon' \leftrightarrow \alpha$, $\gamma'  \leftrightarrow \beta+1$, $\delta'  \leftrightarrow \gamma+1$, $\alpha'  \leftrightarrow \mu+v$, $q' \leftrightarrow \mu$.}
\begin{equation}
   H(\alpha, \beta, \xi) = \sum_{n=0}^{\infty} v_n(\alpha, \beta) \xi^n,
\label{Heunpowerseries}
\end{equation}
with the radius of convergence $|\xi|<1$ given by Poincar\'e-Perron theorem \cite{ronveaux}. The above power series is supplemented with the three-term recurrence relation
\begin{equation}
  A_n v_n=B_n v_{n-1} + C_n v_{n-2}
\label{recurrencerelation}
\end{equation}
with initial conditions $v_0=1$, $v_{-1}=0$ and parameters
$$A_n=1+\frac{\beta}{n}$$
$$B_n=1+\frac{1}{n} \left(\beta+\gamma-\alpha -1 \right)+\frac{1}{n^2} \left( \eta - \frac{1}{2} \left(\beta + \gamma - \alpha \right) + \frac{ \beta}{2} ( \gamma - \alpha)\right)$$
$$C_n=\frac{\alpha }{n}+\frac{\alpha }{n^2} \left( \frac{\delta}{\alpha} + \frac{\beta + \gamma}{2} -1 \right).$$

The solution to Heun's differential equation is given in terms of the infinite power series with the coefficients determined by the above three-term recurrence relation. Unfortunately, it is not possible to find the explicit formula for $v_n$ solving the recurrence relation in Eq. (\ref{recurrencerelation}). Instead the solution may be provided in terms of the holonomic sequence\footnote{A sequence is holonomic if its generating function solves a linear ordinary differential equation with polynomial coefficients \cite{noble2011}.}. For a fixed $\alpha$ parameter, $v_n(\alpha, \beta)$ are polynomials in $\beta$ with their degree increasing with $n$. 

\begin{figure}
\centering
\includegraphics[width=0.85\textwidth]{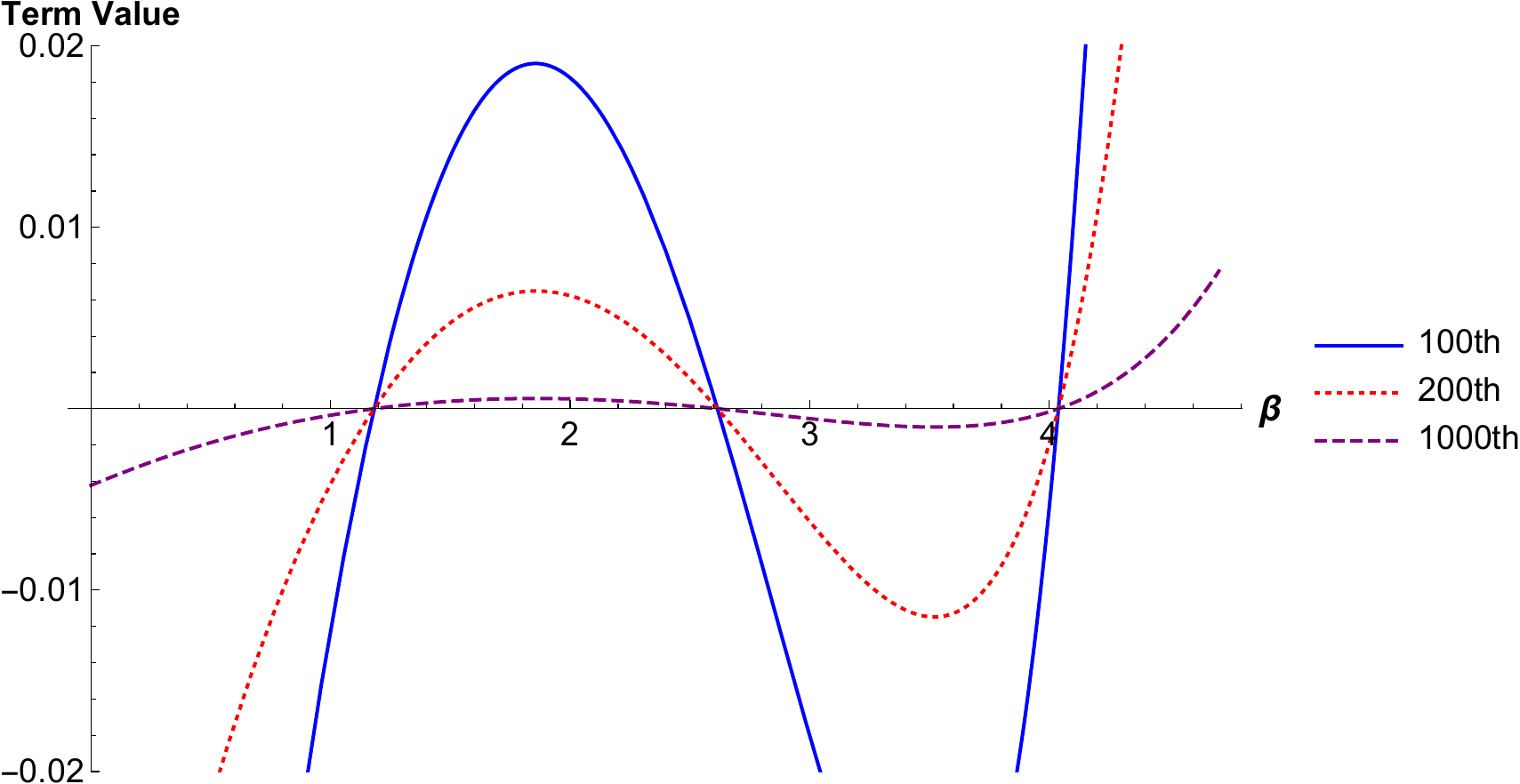}
\caption{$v_n(\beta)$ polynomials for the even parity wavefunctions and parameters $\alpha=-12.229$ and $n=100$, $200$ and $1000$. According to the claim in section \ref{sub:quantisationcondition} the roots of this polynomial for large $n$ correspond to the quantised energy eigenvalues.}
\label{evenquantisationconditionplot}
\end{figure}

Examples of these polynomials are displayed in Fig. \ref{evenquantisationconditionplot} for the even wavefunctions. This figure shows a rather surprising fact that the different degree polynomials in $\beta$ have their roots for almost exactly the same $\beta$ values. This is crucial as it means that if $\beta$ is chosen such that the corresponding $N^{th}$ order polynomial attains $0$, all higher order polynomials $n>N$ will be "very close" to $0$ too, "very close" being further quantified as $\mathcal{O}(1/N)$.

The infinite power series given in Eq. (\ref{Heunpowerseries}) may be terminated to a polynomial of degree $N$ if both $C_{N+2}=0$ and $v_{N+1}=0$ conditions are simultaneously satisfied \cite{downing2013,ronveaux}. In such case the system becomes quasi-exactly-solvable and a subset of its eigenvalues may be found explicitly. However, imposing two equations on one free parameter $\beta$ in the model implies that the other equation must constrain the value of $\alpha$ which represents well location and depth of the potential. This means that, for a fixed well location, terminating to a polynomial approach will be permissible only for selected values of its depth and vice-versa. Unfortunately the above-mentioned constraints did not result in a choice of physically relevant parameters and only one eigenvalue per choice of $\alpha$ may be found \cite{downing2013}. This effectively precludes the calculation of all overlap integrals between the bound-states and the arbitrary initial wavepacket placed in the hyperbolic double-well.

Furthermore, following the similar procedure as above but using the exchange of variables $\zeta=\tanh{(x/d)}$ (with $-1< \zeta < 1$ as $-\infty<z<+\infty$) it may be found \cite{downing2013} that 
\begin{equation}
\psi_{odd}(\zeta)=\zeta \left(1-\zeta^2\right)^{\beta/2} e^{-\frac{\alpha}{2} \zeta^2} H(-\alpha,-\gamma,\beta,-\delta,\eta+\frac{\alpha^2}{4},\zeta^2)
\label{wavefunctions_odd}
\end{equation}
where $H(-\alpha,-\gamma,\beta,-\delta,\eta+\frac{\alpha^2}{4},\zeta^2)$ is again a Heun's function of which coefficients $v'_n$ may be found using three-term recurrence relation Eq. (\ref{recurrencerelation}), only mapping $\alpha \rightarrow - \alpha$, $\beta \rightarrow -\gamma$, $\gamma \rightarrow \beta$, $\delta \rightarrow - \delta$, $\eta \rightarrow \eta+\frac{\alpha^2}{4}$. Note, that in contrast to $x \rightarrow \xi$, the $x \rightarrow \zeta$ transformation intrinsically constrains the wavefunctions to be odd in $x$-space.

\subsection{Quantisation condition \label{sub:quantisationcondition}}

However, it is evident that Schr\"odinger's equation should provide us with the solution for a full-range of depths and well locations of the potential. Instead of working in a quasi-exactly-solvable framework and terminating the Heun power series to a polynomial, we propose an alternative approach. We suggest that the entire eigenspectrum may be found by ensuring that the infinite power series converges to $0$ sufficiently quickly such that the wavefunctions are still square integrable. This possibility stems from the asymptotic (discarding $1/n^2$ terms) behaviour of the holonomic sequence $v_n$: it may be found empirically that the values of its terms for large $n$ significantly depend on the energy quantisation parameter $\beta$. 

We propose that, for a given parameter $\alpha$ of the potential, the admissible values of an energy quantisation parameter $\beta=\beta_{crit}$ (with $\beta>0$) correspond to the roots of the polynomial $v_n (\beta_{crit})$ for large value of $n$ (or more strictly as $n \rightarrow \infty$). Thus, in practice, the problem of finding the allowed energies in the hyperbolic well problem boils down to finding the roots of a certain (usually) high degree polynomial in $\beta$. Based on the above claim we find the energy eigenvalues via numerical root-finding methods. 
This quantisation criterion forms a more efficient alternative to a typically used condition based on Wronskians \cite{xie2012new,hartmann2014,fernandez2011wronskian} and is similar to the one achieved in \cite{hudak1985exact} on different grounds. Furthermore, whenever $\alpha$ is chosen such that the infinite power series terminates, it is clear from Table \ref{tab:comparison} that the eigenvalues found using the above-mentioned claim are arbitrarily close to the ones given by the explicit analytical formula.

\begin{table}
    \centering
    \begin{tabular}{ |l|l|l| }
    \hline
    \multirow{2}{*}[-0.3ex]{\hspace{9.2ex} $\alpha$} & \multicolumn{2}{ |c| }{Eigenvalues} \\
    \cline{2-3}
     &  \multicolumn{1}{ |c| }{Analytic} & \multicolumn{1}{ |c| }{Quantisation} \\
    \hline
    $-12.2300554754797689$ & $2.61502773773988446614$ & $2.61502773773988446614$ \\ 
    \hline
    $-24.4098065308194893$ & $8.70490326540974469239$ & $8.70490326540974469239$ \\ 
    \hline
    \end{tabular}
\caption{Comparison of the eigenenergies obtained using the quantisation condition in section \ref{sub:quantisationcondition} and analytic formula obtained by \cite{downing2013}. The precision displayed corresponds to 20 significant figures.}
\label{tab:comparison}
\end{table}

In other words, we propose that the sequence of the power series coefficients $\{v_n (\alpha, \beta_{crit})\}$ in Eq.~(\ref{Heunpowerseries}) decreases with $n$ quickly enough for allowed values of $\beta$ such that the wavefunction is square-integrable in the range $\xi=0$ to $\xi=1$. In a typical setup this is equivalent to demanding that $\psi_{\rm even}(x \rightarrow \pm \infty) \rightarrow 0$  which corresponds to $\psi_{\rm even}(\xi \rightarrow 0) \rightarrow 0$. Interestingly, in the present case, $\psi_{\rm even}(\xi \rightarrow 0)$ requirement is trivial. On the other hand the $\xi \rightarrow 1$ limit 
lies just "on the edge" of the radius of convergence. 
The $\xi \rightarrow 1$ (corresponding to $x=0$) limit may be investigated using Abel's theorem \cite{abel1826memoire}. The value of the power series as $\xi \rightarrow 1$ should approach $\sum_n^{\infty} v_n (\alpha,\beta)$ provided that $\sum_n^{\infty} v_n (\alpha,\beta)$ converges, which requires choice of $\beta$ (for a fixed $\alpha$) such that $v_n(\beta) < 1/n$ for large $n$ (by direct comparison test). However, it should be noted that convergence of $\psi_{\rm even}(\xi)$ at $\xi=1$ is by no means a \textit{sufficient} criterion for wavefunction square-integrability. Unfortunately, as $v_n$ coefficients are not given by an explicit formula, it seems to be burdensome to evaluate the square-integrability constraint directly to find the admissible values of $\beta$ - especially that the asymptotic (large $n$) behaviour of terms $v_n$ as functions of its parameters, to our best knowledge, is not well-understood \cite{ronveaux}. Therefore, we rely on the indirect arguments presented below. 

\subsubsection{Argument based on the recurrence relation \label{recurrencerelationargument}} 
To motivate the above-stated claim we propose the following  argument based on the recurrence relation (\ref{recurrencerelation}). Consider the particular term $n=N$ of Eq.~(\ref{recurrencerelation}). By $\beta^{crit}_{n}$ we will denote such value of $\beta$, for which the $v_n(\alpha, \beta^{crit}_{n})=0$. 
Suppose that $v_{N-1} (\alpha,\beta^{crit}_{N-1})=0$. Then, using Eq.~(\ref{recurrencerelation}), we obtain 
\begin{equation*}
 \left(1+\frac{\beta^{crit}_{N-1}}{N}\right) v_N = \frac{\alpha}{N^2}\left(\frac{\delta(\alpha,\beta^{crit}_{N-1})}{\alpha}+\frac{\beta^{crit}_{N-1}+\gamma}{2}+N-1\right) v_{N-2}.
\end{equation*}
Evaluating the right-hand-side for large $N$ (i.e when $\alpha/N \gtrsim \alpha\left(\alpha+2\beta^{crit}_{N-1}-5\right)/(4N^2)$ which corresponds to $N \gtrsim |\alpha|$) we arrive at
\begin{equation}
  v_N(\alpha,\beta^{crit}_{N-1}) = \frac{\alpha}{N+\beta^{crit}_{N-1}} v_{N-2}(\alpha,\beta^{crit}_{N-1})  .
\end{equation}
Note that $-1<\alpha/\left(N+\beta^{crit}_{N-1}\right)<0$ provided that $\alpha<0$ and $\beta < |\alpha|$ which is indeed fulfilled as for the normalisable solution to Schrodinger equation we require $E>\min\limits_{x} V(x)$. 

As it was stated in the previous section, the necessary criterion for square-integrability is that the sequence $\{v_n\}$ must converge to $0$ for large $n$ quicker than $1/n$ for large $n$. Therefore, we have that: $v_{N-1}(\alpha,\beta^{crit}_{N-1})=0$ (by assumption) and $v_N(\alpha,\beta^{crit}_{N}) \approx 0 + \mathcal{O}\left(\frac{\alpha}{N}\right)$ with $\alpha/N \ll 1$ hence mimicking the termination of the power series to a polynomial through zeroing two subsequent-terms in the three-term recurrence relation \cite{fiziev2009}. Note that as in principle we can make $N$ to be arbitrarily large, the error associated with the non-exact truncation of the power series can be made arbitrarily small.

Furthermore, it may be readily seen that for $\beta$ chosen such that the polynomial $v_{N-1}(\beta)=0$, all polynomials for $n \geq N-1$ can be factorised into 
\begin{equation*}
v_n(\beta)=q_n(\beta) \left( \frac{\alpha}{N+\beta^{crit}_{N-1}} v_{N-2}(\beta)\right).
\end{equation*}
where $q_n(\beta)$ is another polynomial. Such factorisation property resembles the result from the theory of quasi-exactly-solvable models due to \cite{bender1996quasi}. However, in contrast to \cite{bender1996quasi}, the zeroing of the "critical polynomial" $v_{N-1}(\beta)$ does not imply that all subsequent terms will vanish but instead they will pick up a very small factor of $\alpha/N \ll 1$.

\subsubsection{Argument based on smoothness of the wavefunction \label{smoothnessargument}}

A more strict argument is given by the smoothness of the wavefunction. Although the $x \rightarrow \xi$ mapping intrinsically constrains the wavefunctions to be even, it is not guaranteed that the wavefunction produced by joining of the two half-space wavefunctions will be 'smooth'. However, it is reasonable to demand that their derivatives should be continuous everywhere \cite{sudbery}, i.e., that $\psi'(x=0)=0$ (corresponding to $\xi=1$). This can be expanded to produce
\begin{equation}
\begin{split}
    \frac{d\psi}{dx}\Bigr|_{x=0}=\left[\frac{d\psi}{d\xi}\frac{d\xi}{dx}\right]\Bigr|_{\xi=1}=\left[-2\xi \sqrt{1-\xi} \frac{d\psi}{d\xi}\right]\Bigr|_{\xi=1}=\\ =\lim_{\xi \rightarrow 1} \sum_{n=0}^{\infty} \left[ v_n(\alpha, \beta) (n+\frac{\beta}{2}) \xi^{n+\beta/2-1} e^{\alpha \xi/2} + v_n(\alpha, \beta)  \frac{\alpha}{2} \xi^{n+\beta/2} e^{\alpha \xi /2} \right] \\  \times \left[-2\xi \sqrt{1-\xi}\right]\Bigr|_{\xi=1} = \sum_{n=0}^{\infty} \left[ v_n(\alpha, \beta) \left(  n+\frac{\beta}{2}\right) e^{\alpha/2}  + v_n(\alpha, \beta)  \frac{\alpha}{2} e^{\alpha/2} \right] \\  \times \left[-2\xi \sqrt{1-\xi}\right]\Bigr|_{\xi=1} = \lim_{\xi \rightarrow 1} \left(-2\xi \sqrt{1-\xi}\right)  \times e^{\alpha/2} \sum_{n=0}^{\infty} v_n (\alpha, \beta) \left[ n + \frac{\alpha+\beta}{2} \right].
\end{split}
\end{equation}

The first term of the expression will always be $0$ but the second term needs to be finite for the product to be $0$. For large $n$ we can omit the constants in the second term which becomes: $\sum_{n=0}^{\infty} v_n (\alpha, \beta) n$. Now this expression will converge if $v_n$ goes to $0$ quicker than \textit{$1/n^2$} (by direct comparison test). In other words, the wavefunction $\psi_{\rm even}(x(\xi))$ will be acceptable only for such $\beta$ and certain constant $c$ that $|v_n|<c/n^2$ for sufficiently\footnote{That is, for all $n>k$ for some fixed value $k$.} large $n$ . This is a stronger condition for large $n$ than previously stated. 

\begin{figure}[ht]
\begin{subfigure}{.5\textwidth}
  \centering
  \includegraphics[width=\linewidth]{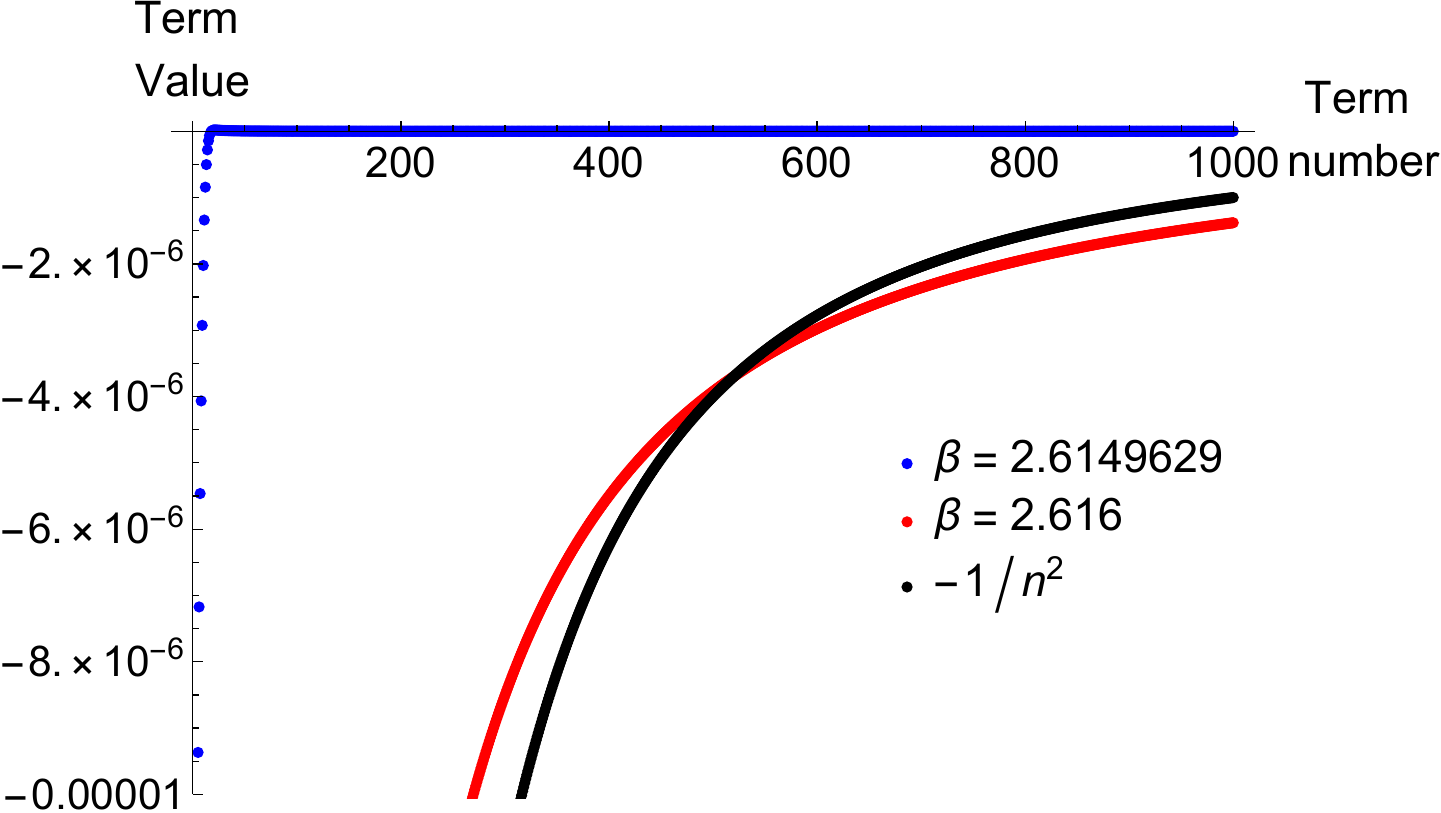}  
\end{subfigure}
\begin{subfigure}{.5\textwidth}
  \centering
  \includegraphics[width=\linewidth]{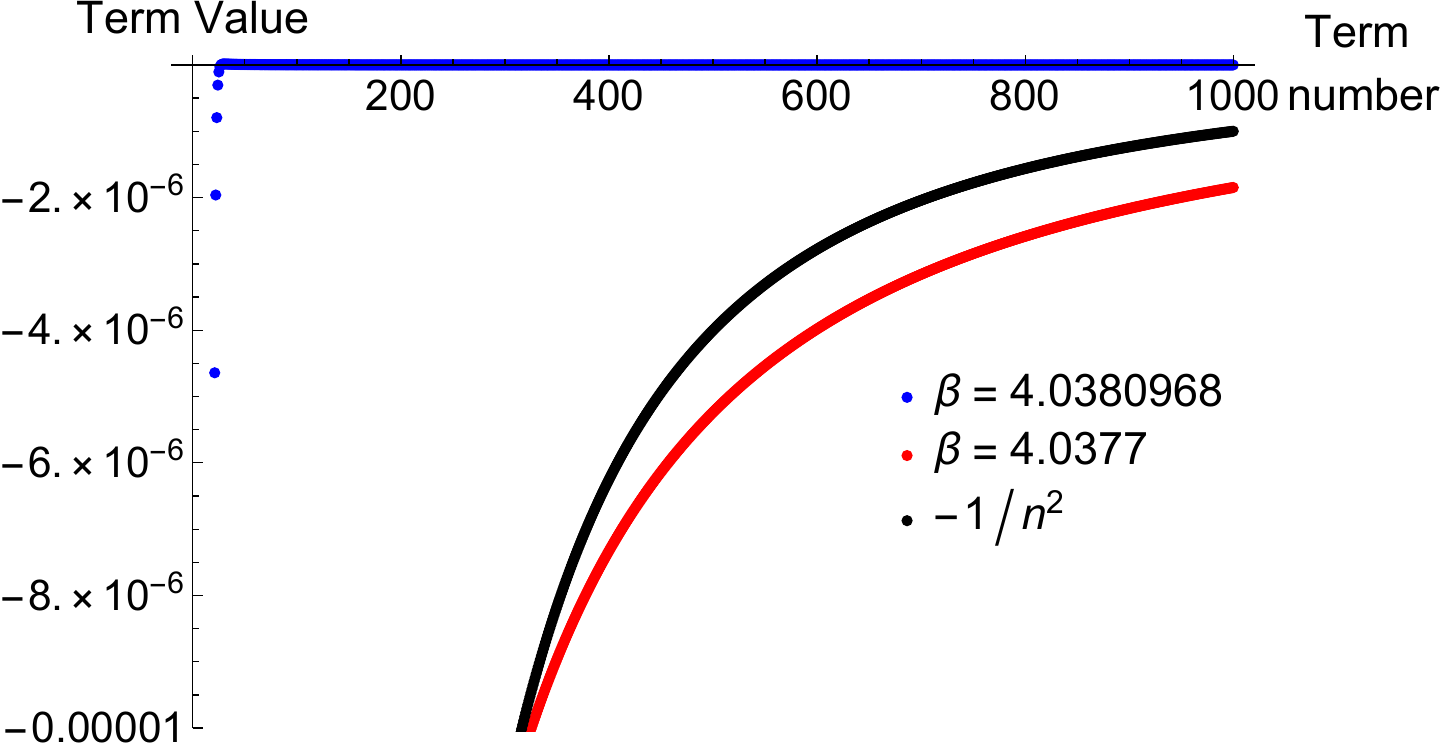}  
\end{subfigure}
\caption{Verification of the $v_n<1/n^2$ upper bound imposed by smoothness. Here $\alpha=-12.229$ and hence relevant $\beta_{crit}=2.614962$ (left panel) and $\beta_{crit}=4.038096$ (right panel).}
\label{fig:convergenceplots}
\end{figure}

Empirical results in Fig. \ref{fig:convergenceplots} indeed confirm that the $v_n(\beta=\beta_{crit})<1/n^2$ with $\beta_{crit}$ given by the quantisation condition from section \ref{sub:quantisationcondition}. Furthermore, values of $\beta$ slightly away from $\beta_{crit}$ do not fulfill this criterion which suggests that the $v_n<1/n^2$ bound is already tight.

Furthermore, the quantisation condition may be rewritten by using a well-established link between continued fractions and three-term recurrence relations. Interestingly, such formulation of the quantisation condition is closely related to a quantisation condition proposed by Manning (1935) (\cite{manning}, p. 137, Eq. (7)) for the potential bearing his name - also a member of a hyperbolic double-well family. The infinite continued-fraction formulation is very convenient for numerical implementations. For details see Appendix A. 

\subsubsection{Wavefunctions and accuracy of a non-exact series truncation}

Here we illustrate that the error attained with the non-exact finite-order truncation of the infinite power series Eq. (\ref{Heunpowerseries}) is negligible. The first clue comes from the factorisation property discussed in section \ref{recurrencerelationargument}: all terms beyond $N-1$ pick up an additional $\alpha/N \ll 1$ factor if $N$ is sufficiently large - and hence should contribute very little to the shape of the Heun function. 

\begin{figure}[ht]
\begin{subfigure}{.5\textwidth}
  \centering
  \includegraphics[width=\linewidth]{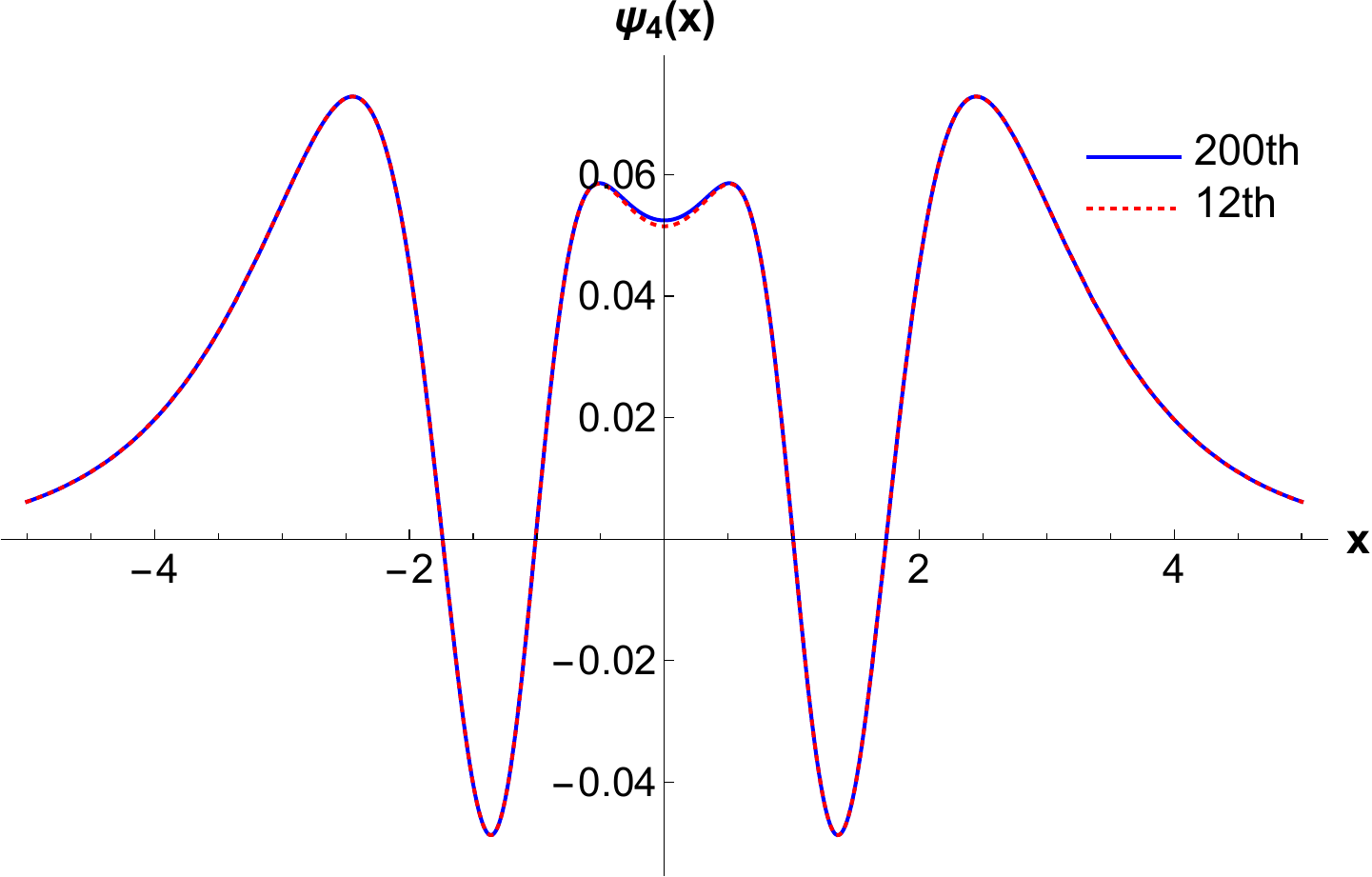}  
  \caption{}
\end{subfigure}
\begin{subfigure}{.5\textwidth}
  \centering
  \includegraphics[width=\linewidth]{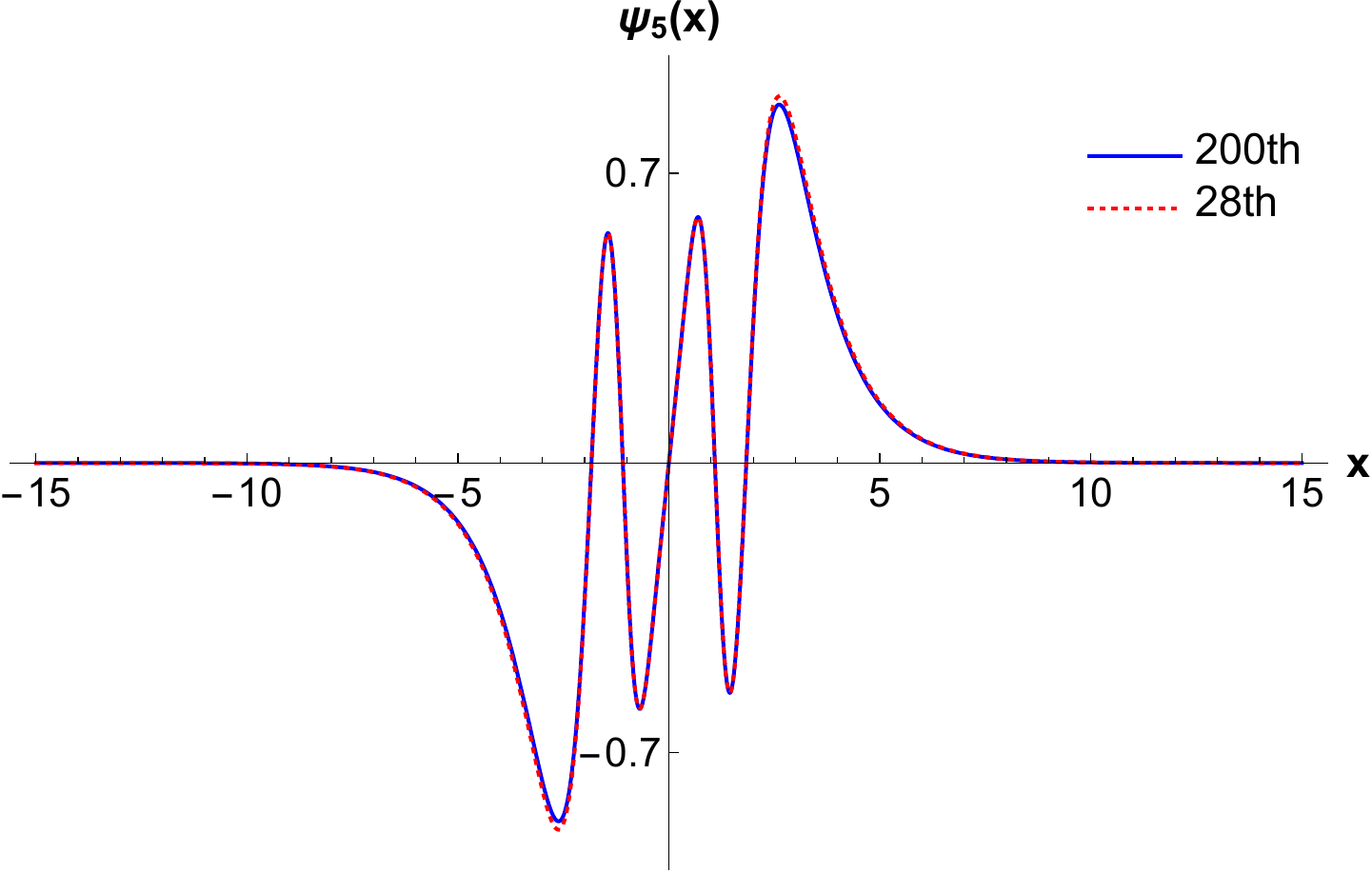}  
  \caption{}
\end{subfigure}
\caption{Comparison of the (not normalised) wavefunctions with the Heun function truncated after different number of terms for parameters $V_0=74.785$, $d=1$. The panels (a) and (b) represent the fourth and fifth excited states.}
\label{fig:truncationplots}
\end{figure}

\begin{figure}[ht]
\begin{subfigure}{0.32\textwidth}
  \centering
  \includegraphics[width=\linewidth]{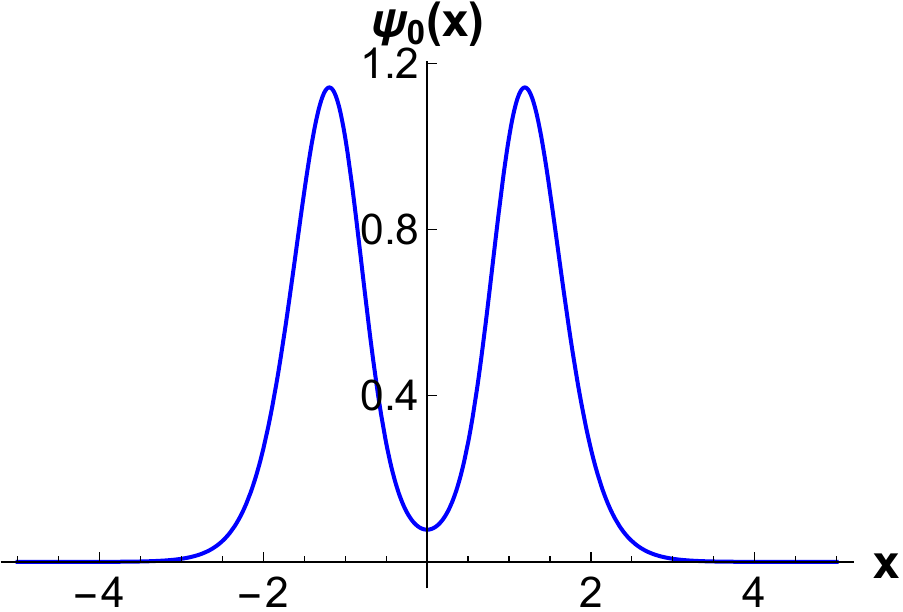}  
\end{subfigure}
\begin{subfigure}{0.32\textwidth}
  \centering
  \includegraphics[width=\linewidth]{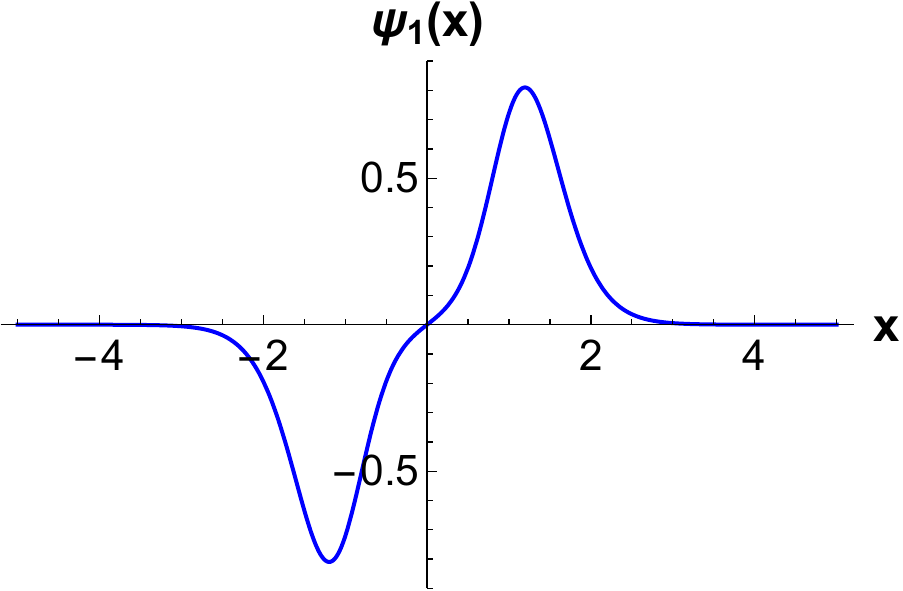}  
\end{subfigure}
\begin{subfigure}{0.32\textwidth}
  \centering
  \includegraphics[width=\linewidth]{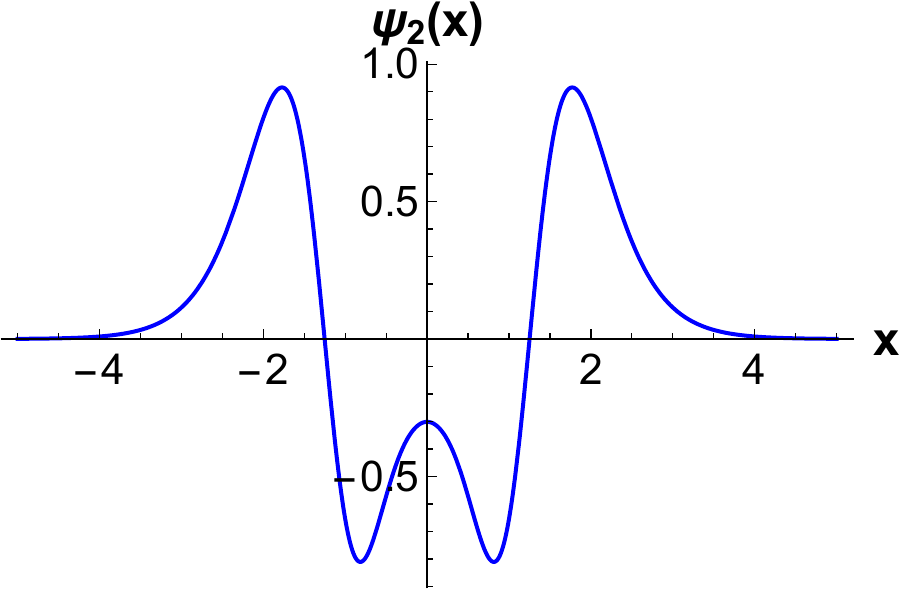}  
\end{subfigure}
\newline
\begin{subfigure}{0.32\textwidth}
  \centering
  \includegraphics[width=\linewidth]{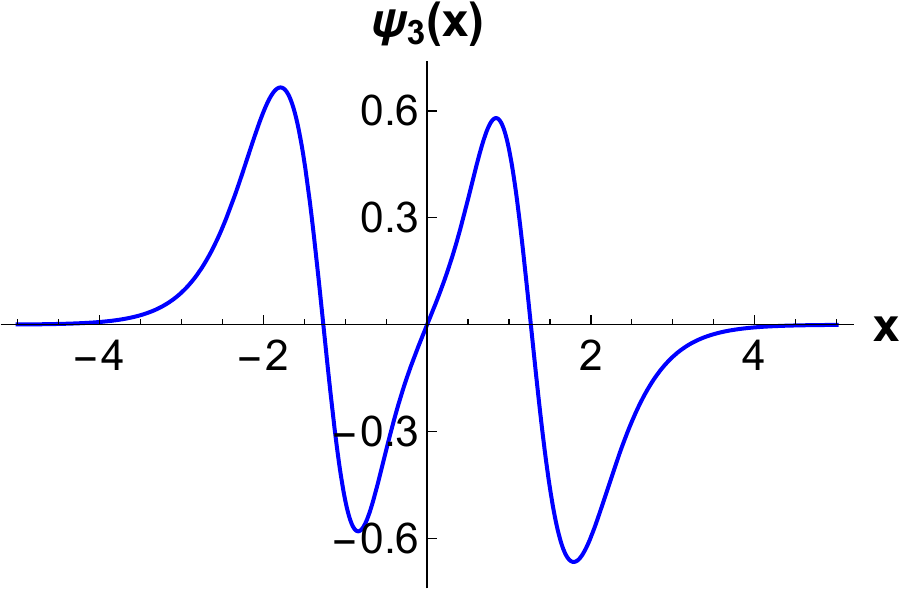}  
\end{subfigure}
\begin{subfigure}{0.32\textwidth}
  \centering
  \includegraphics[width=\linewidth]{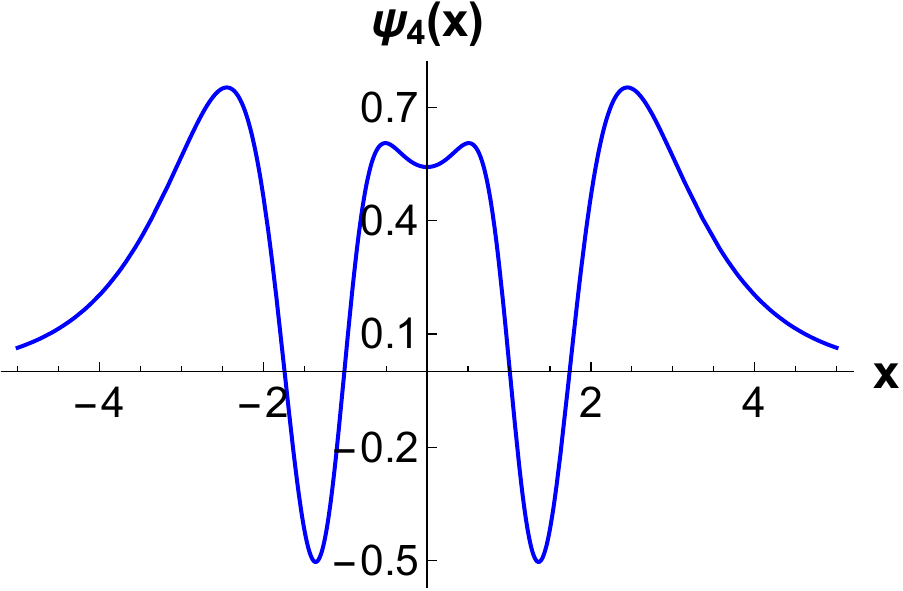}
\end{subfigure}
\begin{subfigure}{0.32\textwidth}
  \centering
  \includegraphics[width=\linewidth]{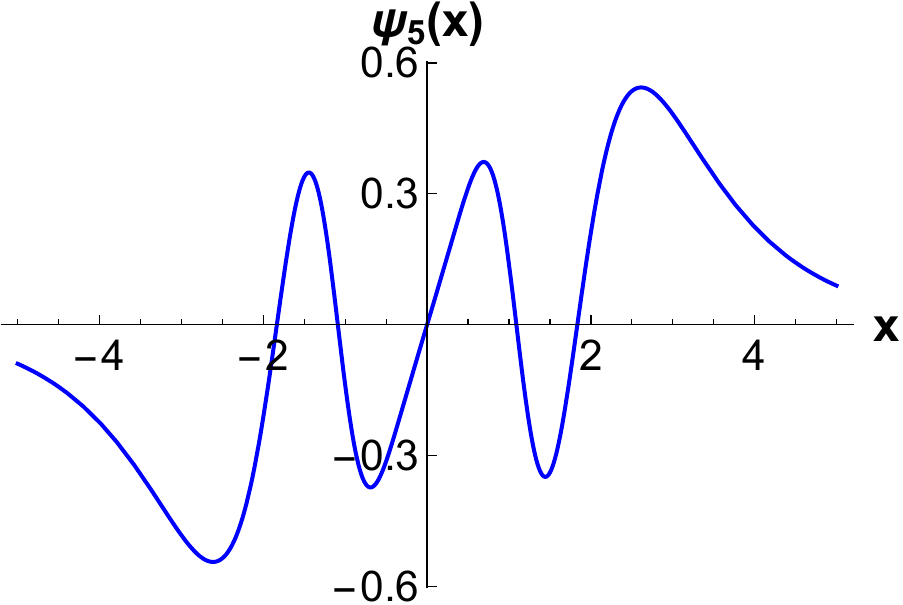}
\end{subfigure}
\caption{Plots of all the bound-state eigenfunctions for parameters $V_0=74.785$, $d=1$. The corresponding eigenenergies (displayed to three significant figures) are: $E=\{-8.153,-8.141,-3.419,-3.298,-0.697,-0.441\}$.}
\label{fig:eigenfunctions}
\end{figure}

On a more practical side, for the parameters chosen ($V_0=74.785$, $d=1 \Rightarrow \alpha=-12.229$), based on Fig. \ref{fig:truncationplots}, it is clear that for terms $n \gtrsim 12$ for the even wavefunctions and $n \gtrsim 28$ for the odd wavefunctions the truncation error may be safely neglected - thus making the non-exact truncation to be extremely accurate and computationally feasible. The set of all bound-state eigenfunctions for parameters $V_0=74.785$, $d=1$ are shown in Fig. \ref{fig:eigenfunctions}. The total number of bound states is discussed in the following section.

\subsection{Number of bound states}
\label{sub:numbBoundStates}
It is not straightforward to predict how many bound states we should expect as functions of $V_0$ and $d$ of the potential, without explicitly invoking the proposed quantisation condition. Here, we instead employ the theoretical lower and upper bounds on the number of bound states $B$. Ref.~\cite{chadan} has given the following upper-bound on the number of bound states in 1D, which was later used in context of hyperbolic-well potentials by \cite{saadandhall} 
\begin{equation*}
B \leq 1+ \sqrt{2} \left[ \int_{-\infty}^{\infty} z^2 V(z) dz \int_{-\infty}^{\infty}V(z) dz \right]^{1/4}.
\label{upperboundboundstates}
\end{equation*}
Furthermore, we recall the well-known theorem stating that the arbitrarily weak potential in one and two dimensions fulfilling $V(x)\leq 0$ for all $x$ and $\int_{-\infty}^{+\infty} V(x) d^n x <0$ ($n=1,2$) will have a bound state (see e.g. \cite{chadan} p. 2) for reference). 
Therefore, we conclude that the number of bound states in a hyperbolic double-well potential is bounded by
\begin{equation}
    1 \leq B \leq 1 + \left({\frac{4}{75}}\left(20+\pi ^2\right)\right)^{1/4} (V_0 d)^{1/2}.
\end{equation}
It may be clearly seen that for the hyperbolic-double-well potential we expect the number of bound states to be finite, with an upper bound growing like $(V_0 d)^{1/2}$. For parameters of the potential $V_0=74.785$ and $d=1$ this results in the number of bound states $1 \leq B \leq 10$ which provides rather tight bounds on the actual number of bound states found from a quantisation condition: $B=6$. 

\subsection{Initial wavepackets, overlap integrals and temporal evolution}
\label{sub:wavepackets}
Having found the eigenfunctions and eigenvalues of the TISE, we now formulate the initial wavepackets to be placed in a hyperbolic double-well. The purely even/odd (when mapped to the $x$-space) wavepackets may be divided in $\xi$- and $\zeta$-spaces respectively. We would like to benefit from the relatively simple forms of the eigenfunctions in $\xi$- and $\zeta$-spaces and to devise the simple purely even/odd initial wavepackets in $\xi$- and $\zeta$-spaces. In this way the intricate\footnote{For example, the even-parity eigenfunctions from Eq. (\ref{wavefunctions_even}) when evaluated in $x$-space become $\psi(x)=(1/cosh^2(x/d))^{\beta/2} e^{\alpha (1/2cosh^2(x/d))} H(\alpha,\beta,\gamma,\delta,\eta,1/cosh^2(x/d))$} overlap integrals from $x$-spaces can be evaluated in $\xi$- and $\zeta$-spaces by introducing the appropriate measures (weight functions) $q(\xi)$ and $Q(\zeta)$ to the integrals. In $\xi$-space, noting that $d \xi /dz=-2\sinh{(z)}/\cosh^3{(z)}$ we obtain
\begin{equation}
    dx=\frac{d}{-2\xi \sqrt{1-\xi}} d\xi=-q(\xi,d) d\xi,
\end{equation}
whereas in $\zeta$-space we have: $d\zeta/dz=1/\cosh(z)^2=1-\tanh(z)^2=1-\zeta^2$ and therefore
\begin{equation}
    dx=\frac{d}{1-\zeta^2} d\zeta = Q(\zeta,d) d\zeta.
\end{equation}

\subsubsection{Delocalised wavepackets}

\begin{figure}
\begin{subfigure}{0.5\textwidth}
  \centering
  \includegraphics[width=\linewidth]{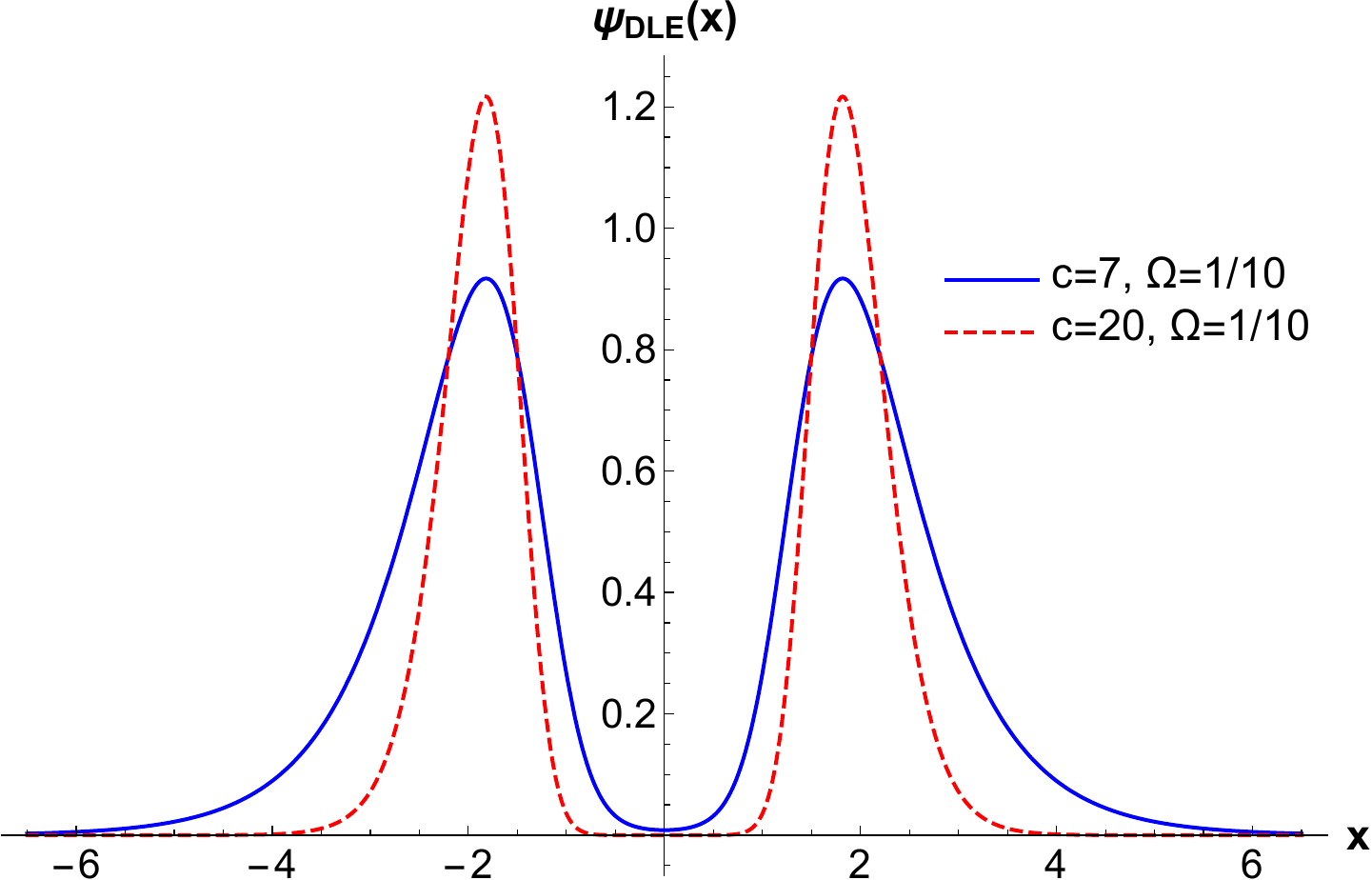}  
  \caption{}
\end{subfigure}
\begin{subfigure}{0.5\textwidth}
  \centering
  \includegraphics[width=\linewidth]{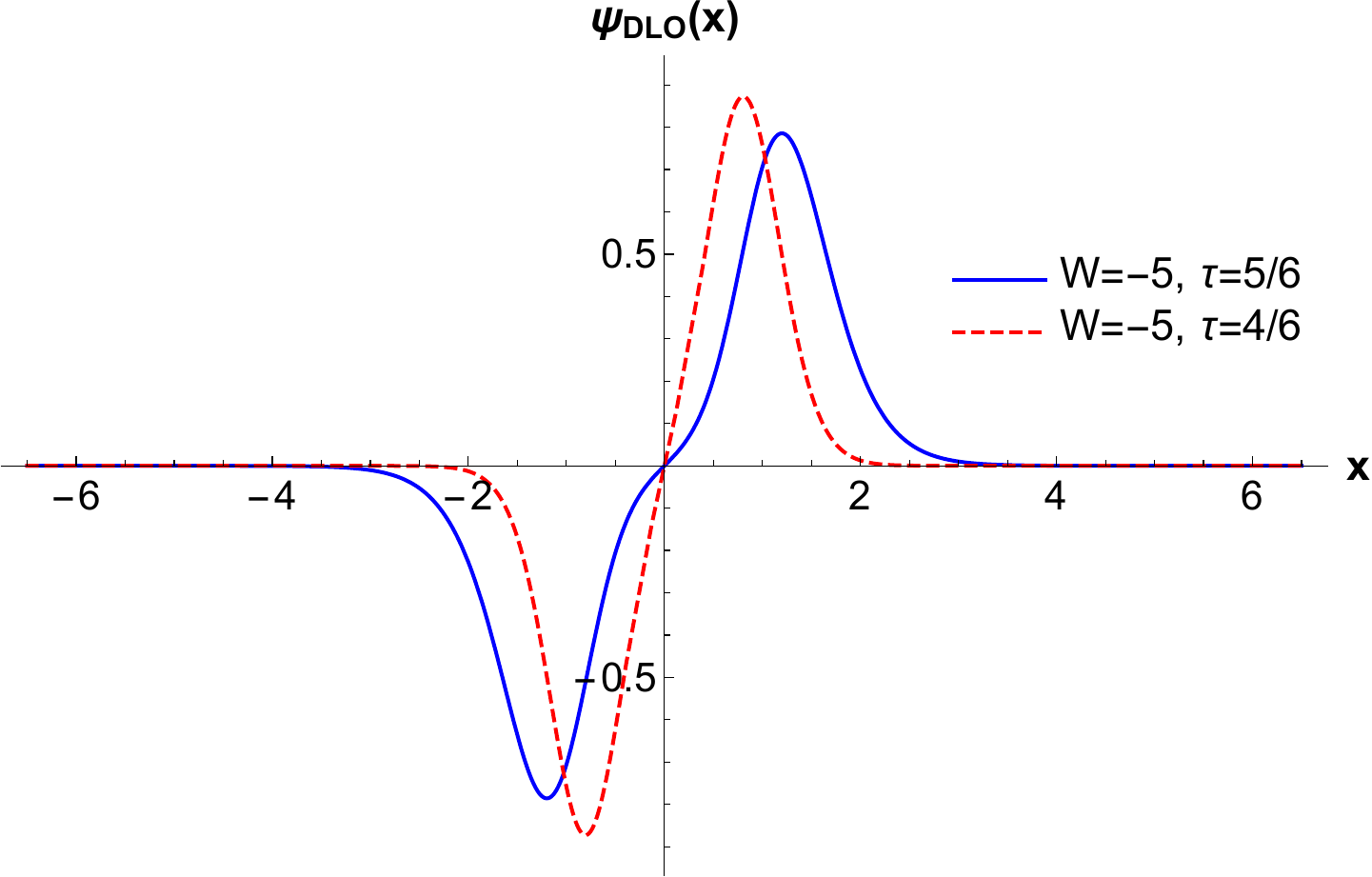}  
  \caption{}
\end{subfigure}
\caption{Graphs of the $\psi_{\rm DLE}(x,W,\tau)$ (left) and $ \psi_{\rm DLO}(x,W,\tau)$ (right). Note that in both cases width and peak location of the wavepackets may be modified independently.}
\label{fig:initialwavepackets}
\end{figure}

Next we propose even (odd) initial wavepackets  $\Psi(\xi,0)=\psi_{\rm DLE}(\xi,c,\Omega)$  $(\psi_{\rm DLO}$ $(\zeta,W,\tau,d)$) such that their widths and peak location can be chosen independently by modifying parameters $c$ and $\Omega$ ($W$ and $\tau$) respectively. Both are properly normalised with regard to a $q(\xi,d)$ ($Q(\zeta,d)$) measure. The plots of $ \psi_{\rm DLE}(\xi,c,\Omega)$ and $ \psi_{\rm DLO}(\zeta,W,\tau)$ are displayed in Fig. \ref{fig:initialwavepackets}(a) and (b) respectively. 

Explicitly, the even-parity initial wavepacket $\Psi(\xi,0)=\psi_{\rm DLE}(\xi,c,\Omega)$ reads
\begin{equation}
    \psi_{\rm DLE}(\xi,c,\Omega)=\frac{ \xi^{c \Omega} e^{-c \xi}}{\sqrt[4]{\pi } \sqrt{\, _1\tilde{F}_1\left(2 c \Omega ;2 c \Omega +\frac{1}{2};-2 c\right)} \sqrt{\Gamma (2 c \Omega )}},
\label{initialwavepacketeven}
\end{equation}
with $0 < \Omega < 1$; $c>0$; $_1\tilde{F_1} (a,b,c)$ denoting regularised confluent hypergeometric function and $\Gamma(u)$ complete gamma function. It fulfills the following properties
\begin{itemize}
 \item The functional form of the even wavepacket represents a constrained $\beta$-distribution in $\xi$-space. The functional form of this distribution is proportional to $\xi^p e^{-c \xi}$ with ($p=c \Omega>-1$ and $c>0$) and. To ensure the finite value of $\psi_{\rm DLE}$ at $\xi=0$ we require $p > 0$. 
    \item $\psi_{\rm DLE}(x,c,\Omega)$ contains two parameters which \textit{independently} specify width and location of the peak. This is a result of a particularly simple relation between the location of a peak of a $\beta$-distribution as a function of parameters $p$ and $c$: $$\left[ \frac{d}{d\xi}\left(\xi^p e^{-c \xi}\right)=0 \leftrightarrow \xi_{peak}=\frac{p}{c} \defeq \Omega \right].$$ Therefore we set $p=\Omega c$ with $\Omega$ solely specifying the location of the peak and $c$ solely specifying the wavepacket width.
\end{itemize}

The odd parity wavepacket is given by 
\begin{equation}
\psi_{\rm DLO}(\zeta,W,\tau,d)=\frac{\sqrt{2} \zeta  e^{-W\zeta ^2} \left(1-\zeta ^2\right)^{\frac{\left(\tau ^2-1\right) \left(2 \tau ^2 W-1\right)}{2 \tau ^2}}}{\sqrt[4]{\pi } \sqrt{d \,
   _1\tilde{F}_1\left(\frac{3}{2};2 W \left(\tau ^2-1\right)+\frac{1}{\tau ^2}+\frac{1}{2};-2 W\right) \Gamma \left(2 W \left(\tau ^2-1\right)+\frac{1}{\tau
   ^2}-1\right)}},
   \label{initialwavepacketodd}
\end{equation}
with $W < \frac{1}{2 \tau^2}$ and $0<\tau<1$ and $\Gamma(u)$ denoting a complete gamma function and $_1\tilde{F_1} (a,b,c)$ regularised confluent hypergeometric function. Its choice has been motivated by the properties stated below:
\begin{itemize}
\setlength\itemsep{0.1em}
    \item The functional form of the odd parity wavepacket is proportional to $\zeta \left( 1-\zeta^2 \right)^P e^{-W \xi^2}$. Such form may be motivated by noting that the odd wavepacket should have zeros at $\zeta=\pm 1$ (corresponding to $x=\pm \infty$) and at $\zeta=0$ (corresponding to $x=0$). Constraint $W < \frac{1}{2 \tau^2}$ stems from demanding finite value of $\psi_{\rm DLO}$ at $\zeta=\pm 1$ (and hence $P>0$) along with noting that $P$ is a decreasing function of $W$ for $-1<\tau<1$ (see bullet point below). 
    \item $\psi_{\rm DLO}(\zeta,W,\tau,d)$ contains two parameters which \textit{independently} specify width $W$ and location of the peak $\tau$. This may be shown by setting $P=\frac{\left(\tau ^2-1\right) \left(2 \tau ^2 W-1\right)}{2 \tau ^2}$.
   
\end{itemize}
For such wavepackets the overlap integrals $\Lambda_n=\int_{0}^{1} \psi_{\rm DLE}^{\ast} (\xi, c, \Omega) \psi_n(\xi) q(\xi,d) d\xi$ (for $n=0,2,4,...$) and $\Lambda_n=\int_{-1}^{1} \psi_{\rm DLO}^{\ast} (\zeta, W, \tau) \psi_n(\zeta) Q(\zeta,d) d\zeta$ (for $n=1,3,5...$) can be calculated in terms of the incomplete gamma functions $\Gamma(a,u)$.
The $x$-space peaks location of the even/odd delocalised wavepacket can be easily retrieved from $\Omega$ or $\tau$ by inverting the $x \rightarrow \xi$ and $x \rightarrow \zeta$ mappings to produce $x=\pm d \cosh^{-1}\left( 1/\sqrt{\Omega} \right)$ or $x=\pm d \tanh^{-1} (\tau)$. 

\subsubsection{Arbitrary wavepackets and a temporal evolution \label{temporalevolutionsection}}

Wavefunctions $\phi(\xi)$ from $\xi$-space and $\phi(\zeta)$ from $\zeta$-space are intrinsically mapped to even and odd $x$-space wavefunctions respectively. To produce arbitrary localised states we note that because functions in $\xi$- and $\zeta$-spaces are orthogonal to each other when evaluated in $x$-space, we can just form the linear combinations of the initial wavepackets from both spaces to get neither purely even nor purely odd initial wavepacket in $x$-space. In such case, for the wavepacket of the form $\Psi_{g}(\xi,\zeta,0)=\cos{(\Delta)} \psi_{\rm DLE}(\xi) + \sin{(\Delta)} \psi_{\rm DLO}(\zeta)$, the overlap integrals become

\begin{equation*}
    \chi_n=\int_{-\infty}^{+\infty} \Psi_{g}(\xi(x),\zeta(x),0) \psi_n(x) dx=
    \begin{cases}
\Lambda_n \cos(\Delta) & \text{for } n=0,2,4 \\
\Lambda_n \sin(\Delta) & \text{for } n=1,3,5
\end{cases}
\end{equation*}
where we have used the fact that $\psi_{\rm DLE}(\xi(x))$ and $\psi_{\rm DLO}(\zeta(x))$ are respectively purely even and odd in an $x$-space. 
Furthermore, it should be noted, that as it is possible to calculate the overlap integrals for arbitrary values of $(c, \Omega)$ and $(W, \tau)$ (subject only to constraints imposed in the previous sections), almost an arbitrary wavepacket in the x-space may be formed by making the linear combinations of the purely even (or odd) wavepackets with the fixed width $c$ (or $W$) and varying peak location $\Omega$ (or $\tau$).

\section{Applications to tunneling dynamics}
\label{sec:applications}

Next we will apply the analytical model developed here to the tunneling dynamics of an electronic wave packet, with focus on non-adiabatic temporal evolution. Our motivation is related to the presence of momentum gates, which have been first identified in the context of strong-field enhanced ionisation in position-momentum phase space using Wigner quasiprobability distributions \cite{takemoto2011}. Momentum gates are lines of approximately constant momentum through which there is a direct intra-molecular quasiprobability flow from one molecular centre to the other. In  \cite{takemoto2011}, they have been attributed to the non-adiabatic effect of a transient electron localisation at one of the wells due to the presence of a strong laser field. Such behaviour was further expounded by \cite{chomet2019} who have shown that the time-dependent field is not a necessary prerequisite for the momentum gates to occur and that the strong quasi-probability transfers may occur through "quantum bridges". These are highly non-classical, cyclic structures that form due to quantum interference.  The aim of this section is to quantify this evolution for different initial wave packets, both in time and phase space. 

\subsection{Temporal evolution of the wavepacket
\label{sec:autocorr}}
Now, we exploit the ability to calculate the overlap integrals in the hyperbolic-double well. This may be done in few steps: 
\begin{enumerate}
    \item We fix the parameters of the hyperbolic-double-well potential and find the eigenvalues $E_n$ (and hence eigenfrequencies $\omega_n$) based on the quantisation condition provided in section \ref{sub:quantisationcondition}.
    \item We find the eigenfunctions given by Eqs. (\ref{wavefunctions_even}) and (\ref{wavefunctions_odd}) for the values of $\beta$ (allowed energies) found previously.
    \item We devise the initial even/odd wavepackets according to Eqs. (\ref{initialwavepacketeven}) and (\ref{initialwavepacketodd}).
    \item We calculate the overlap integrals in terms of incomplete gamma functions $\Gamma(a,u)$ as detailed in section \ref{temporalevolutionsection}.
\end{enumerate}

The results of applying this procedure are presented here for two different sets of parameters, each corresponding to a different limit behaviour. 
The time evolution of a wavepacket can be inferred using an autocorrelation function 
\begin{equation*}
    a(t)=\int_{-\infty}^{+\infty} \Psi^\ast (x,t) \Psi(x,0) dx = \sum_{n} |\Lambda_n|^2 \exp\left(\frac{iE_nt}{\hbar}\right) ,
    \label{autocorrdefinition}
\end{equation*}
where  $\Lambda_n$ are the overlap integrals defined in Eq.~(\ref{eq:overlapintegrals}). Therefore
\begin{equation}
|a(t)|^2=\sum_n \sum_m |\Lambda_n|^2 |\Lambda_m|^2 \exp \left( i (E_m-E_n) t/\hbar\right). 
\label{autocorr}
\end{equation}
Thus, any time dependence of $|a(t)|^2$ will stem from the differences in eigenenergies. Note that if $|\Lambda_n| \neq 0$ and $|\Lambda_m| \neq 0$ only for one pair of $n$ and $m$ with $n \neq m$ then $|a(t)|^2$ will oscillate with a single frequency. Otherwise, the time evolution will be more involved. 

\begin{figure}[ht]
\centering
\begin{subfigure}{0.45\textwidth}
  \centering
  \includegraphics[width=\linewidth]{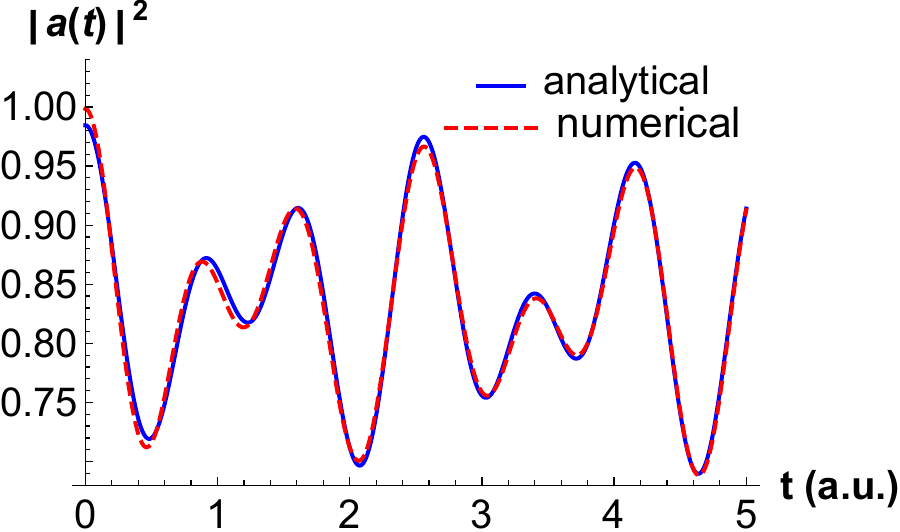}  
  \caption{}
\end{subfigure}
\begin{subfigure}{0.45\textwidth}
  \centering
  \includegraphics[width=\linewidth]{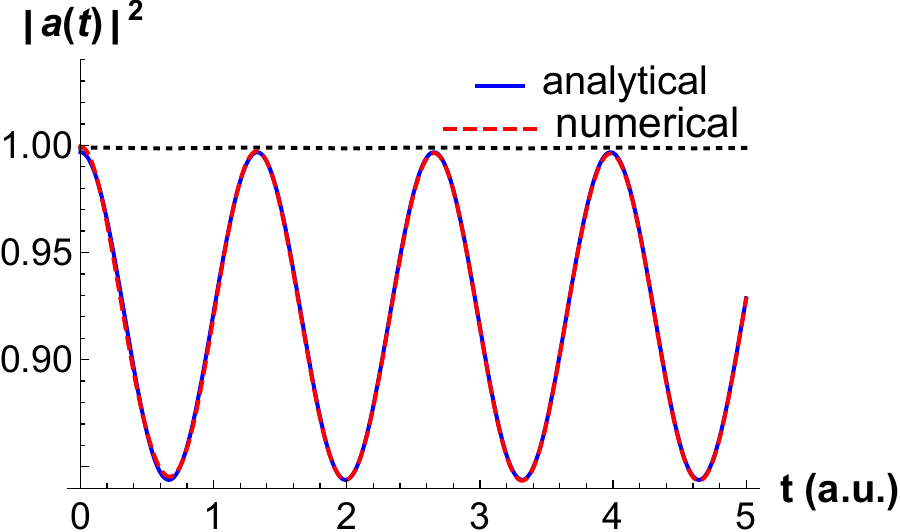}  
  \caption{}
\end{subfigure}
\caption{Comparison of  behaviour of  $|a(t)|^2$ calculated using the analytical method (blue, solid line) described in Sec.~\ref{sub:quantisationcondition} and the numerical method (red, dashed line) in \cite{chomet2019}. In panel (a) $c=4$, $\Omega=1/4$ and in panel (b) $c=7$, $\Omega=1/4$. The horizontal, black, dotted line in panel (b) corresponds to parameters $c=7, \Omega=3/10$, for which the initial wave packet very closely resembles the $\psi_0(x)$ eigenstate, hence having only a minute time dependence. The parameters of the potential (Eq. \ref{potential_generic}) are $V_0=74.785$ a.u., $d=1$ a.u. which corresponds to internuclear distance $R \approx 2.28$ a.u..}
\label{fig:autocorr_pure_even_wavepacket}
\end{figure}

First we devise an even-parity, delocalised wavepacket.
The absolute value squared of the autocorrelation function $|a(t)|^2$ is displayed in Fig. \ref{fig:autocorr_pure_even_wavepacket} for wavepackets of different widths. They have been computed analytically using the method developed above, and numerically using the method in \cite{chomet2019}. The agreement is excellent, with the analytical and numerical curves being practically indistinguishable and the temporal behaviours depending critically on the width.

In Fig.~\ref{fig:autocorr_pure_even_wavepacket}(a) this behaviour is quite intricate with two main frequencies: $\omega_{20}=4.73~$ a.u. and $\omega_{40}=7.46~$ a.u.. This is due to the coupling of $n=0$ with $n=2$, and $n=0$ with $n=4$ eigenstates (as $\Lambda_2$ and $\Lambda_4$ are small the $n=2$ with $n=4$ coupling may be safely neglected). In contrast, in Fig.~\ref{fig:autocorr_pure_even_wavepacket}(b), one can identify a single frequency for $|a(t)|^2$, namely $\omega_{20}=4.73~$ a.u., which corresponds to only one \textit{pair} of states with non-vanishing overlap integrals: $\Lambda_0$ and $\Lambda_2$. Finally, the straight horizontal line in panel (b) corresponds to an initial wavepacket being very close to an eigenstate. As expected, this leads to a constant  $|a(t)|^2$ within the precision used here. Minor discrepancies between the analytical and numerical results are related to the former not including overlaps with scattering states. 

\begin{figure}[ht]
\begin{subfigure}{0.5\textwidth}
  \centering
  \includegraphics[width=0.95\linewidth]{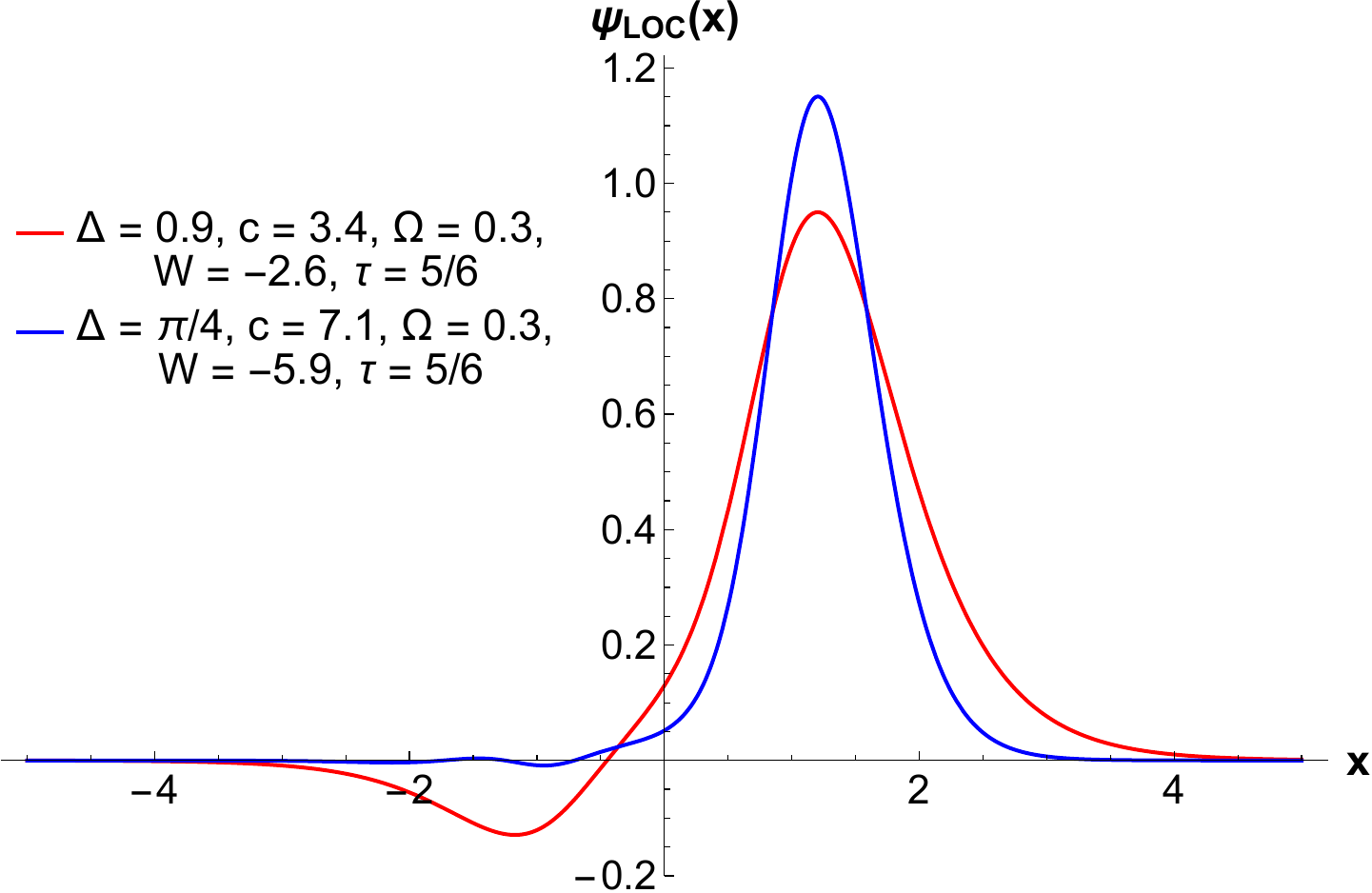}  
  \caption{}
\end{subfigure}
\begin{subfigure}{0.5\textwidth}
  \centering
 \includegraphics[width=0.95\linewidth]{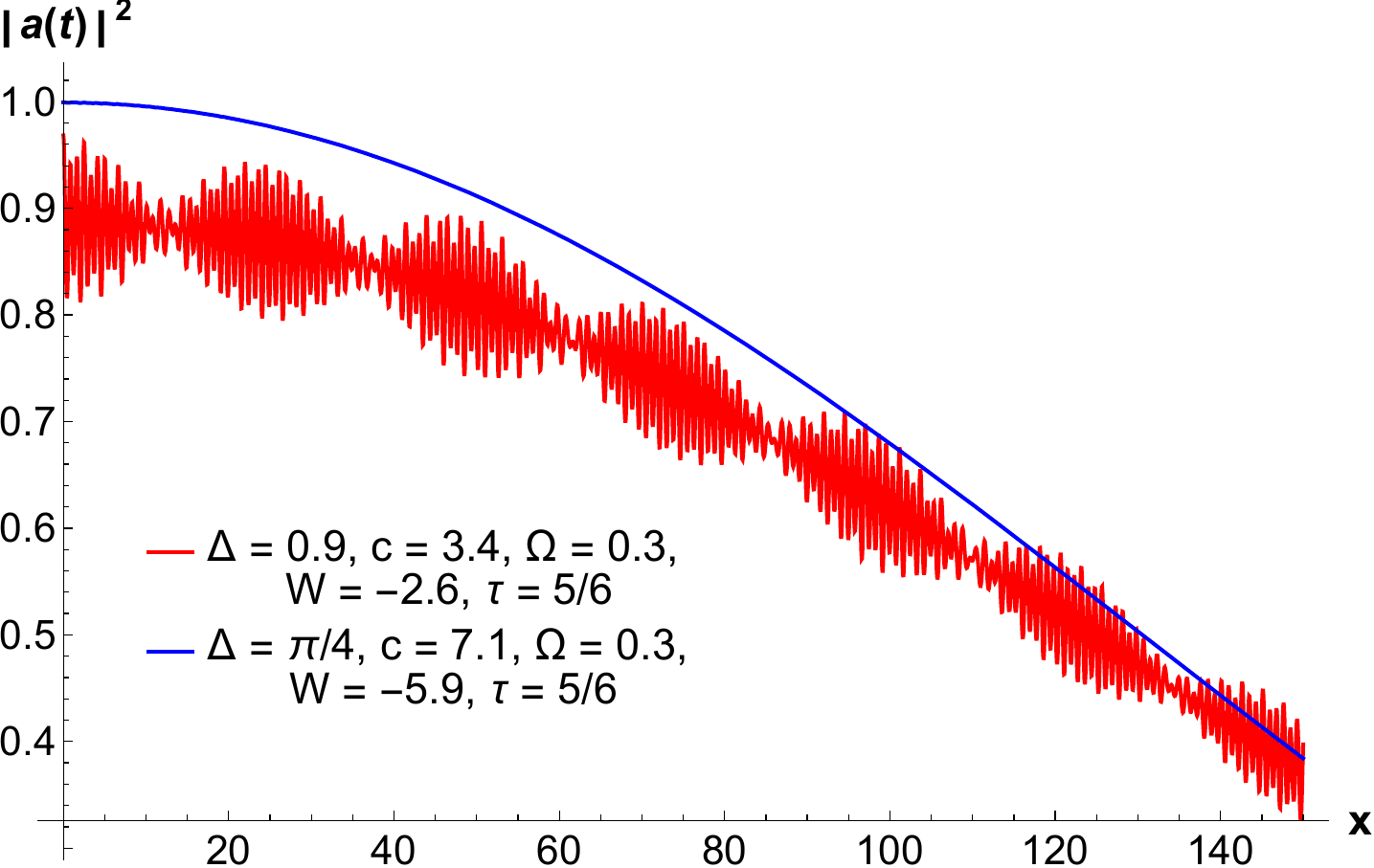}  
  \caption{}
\end{subfigure}
\caption{$\Psi(x,0)$ wavepackets (left) and $|a(t)|^2$ function (right) for two different sets of the $(\Delta, c, \Omega, W, \tau)$ parameters. The parameters of the potential used are $V_0=74.785$, $d=1$.}
\label{fig:wavepacket_localised}
\end{figure}

This critical behaviour is also observed for initially localised wavepackets such as those presented in Fig.~\ref{fig:wavepacket_localised}(a). The corresponding values of  $|a(t)|^2$  are displayed in Fig.~\ref{fig:wavepacket_localised}(b) and are quite distinct. The blue curves in both panels illustrate the scenario in which only the overlap integrals $\Lambda_0$ and $\Lambda_1$ are non-vanishing. In contrast, the red curves show a slightly different wavepacket which gives rise to several different frequencies in the modulus squared of the autocorrelation function. One should note that for localised wave packets there are contributions from both even and odd eigenstates, which may lead to pulsated high-frequency oscillations enveloped by a slow oscillation.  

We produce the localised wavepackets by making different linear combinations of initial wavepackets: $\psi_{\rm DLE}(\xi)$ and $\psi_{\rm DLO}(\zeta)$. Various modes of behaviour are displayed in Fig. \ref{fig:wavepacket_localised}. Note that as the initial wavepackets (for the choices of parameters made) strongly overlap with the ground/first-excited states any short-scale oscillations in the $|a(t)|^2$ are enveloped with the long-scale oscillation of a period $T = 2\pi/(E_1-E_0) \approx 520 (a.u.)$. 

\subsection{Phase-space dynamics}
\label{sec:Wigner}

Next we will investigate the wave packet's phase space evolution, with emphasis on the quantum bridges and their periodic motion. For that purpose, we will employ Wigner quasiprobability distributions. They are given by
\begin{equation}
W(x,p,t)= \frac{1}{\pi} \int_{-\infty}^{\infty}d\mu\Psi^{*}(x+\mu,t)\Psi(x-\mu,t)e^{2ip\mu}, 
\label{eq:Wigner}
\end{equation}
where the position and momentum coordinates are represented by \(x\) and \(p\), respectively. Eq.~(\ref{eq:Wigner}) provides momentum and position resolution, within the constraints posed by the uncertainty principle. It also leads to the probability density in position or momentum space if integrated over the momentum or position coordinates, respectively. One should note, however, that Eq.~(\ref{eq:Wigner}) can be negative making it is a quasiprobability distribution. For more details on quantum systems in phase space see, e.g., \cite{schleich2011quantum}. 
In the analytical model $W(x,p,t)$ may be calculated by numerical integration of Eq.~(\ref{eq:Wigner}), with the temporal evolution of the wavepacket $\Psi(x,t)$ given by Eq.~(\ref{eq:wpcoherent}). The wavepacket, the eigenenergies and the eigenfunctions are calculated analytically as discussed in the previous sections.  

\begin{figure}[ht]
    \centering
    \includegraphics[width=0.9\linewidth]{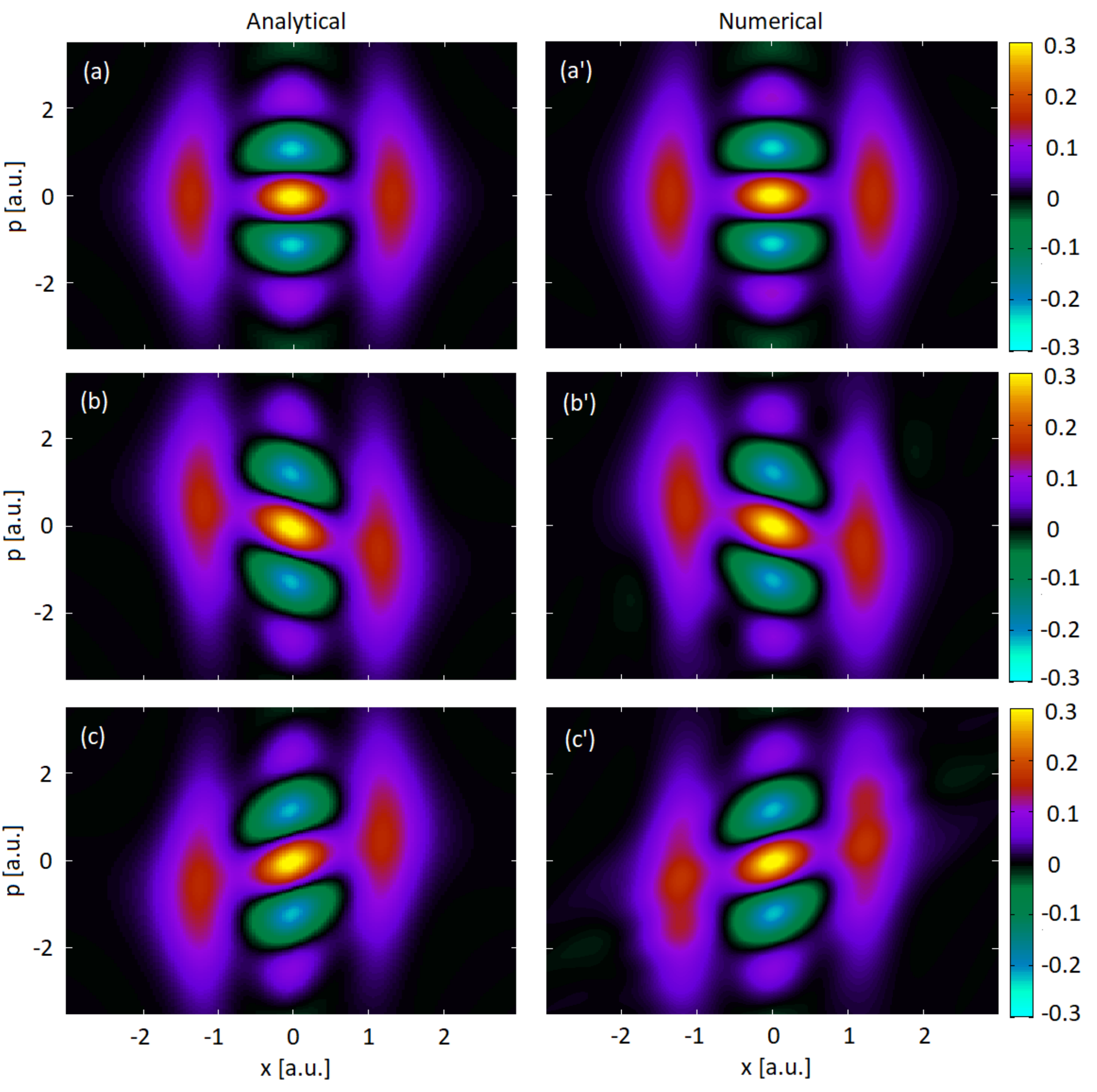}
    \caption{Comparison of Wigner quasiprobability distributions using the same parameters as in Fig.~\ref{fig:autocorr_pure_even_wavepacket}(b) ($c=7, \Omega=1/4$, $R=2.28$) computed analytically (left panels) and numerically (right panels) for the times (a)  $t=0$, (b) $t=0.4$ and (c)  $t=1.0$.}
    \label{fig:wignerPure}
\end{figure}

\begin{figure}[ht]
    \centering
    \includegraphics[width=0.95\linewidth]{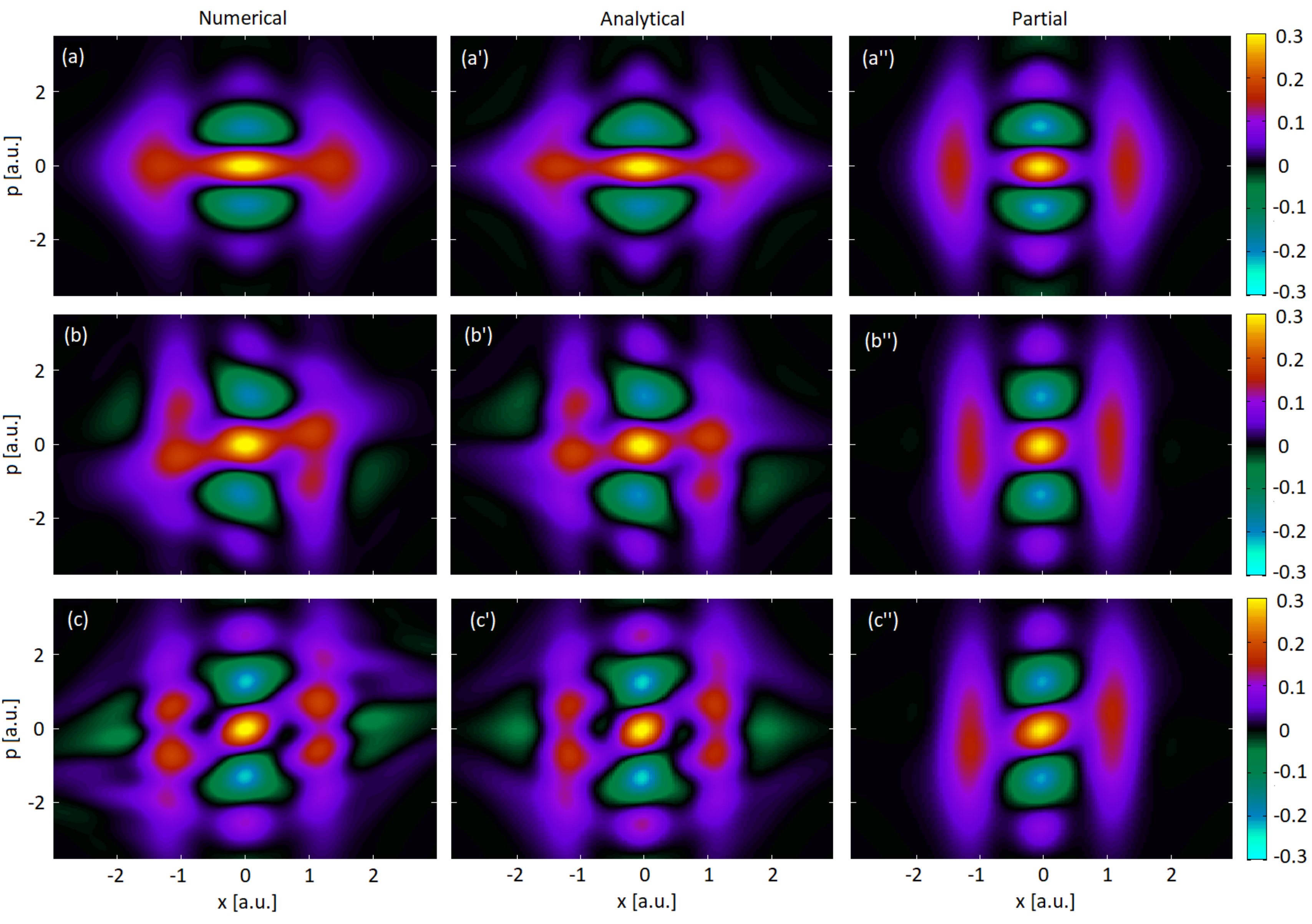}
    \caption{Comparison of Wigner quasiprobability distributions using the same parameters as in Fig.~\ref{fig:autocorr_pure_even_wavepacket}(b) ($c=4, \Omega=1/4$, $R=2.28$) computed numerically (left panels) and analytically (middle panels) for the times (a), (a') and (a'')  $t=0$; (b), (b') and (b'') $t=0.7$; and (c), (c') and (c'') $t=2.1$. In the rightmost panels the Wigner quasiprobability distribution is computed using the analytical model for the partial coherent superposition in Eq.~(\ref{eq:partialphi}).}
    \label{fig:wignerMixed}
\end{figure}

Throughout, we will focus on the scenario for which the quantum bridges are strong, namely initially delocalised wave packets and intermediate internuclear separations. The results comparing the present analytical model and the numerical results in \cite{chomet2019} are displayed in Figs.~\ref{fig:wignerPure} and \ref{fig:wignerMixed}. Fig.~\ref{fig:wignerPure} corresponds to an initial wave packet leading to a single oscillation frequency in the autocorrelation function, while in Fig.~\ref{fig:wignerMixed} a more involved scenario with superimposed oscillations is explored. Overall, the agreement between the numerical and analytical results is excellent, which once more shows that the present model is reliable and, in contrast to the numerical approach in \cite{chomet2019}, can be used to determine the temporal evolution of the quantum bridges exactly. 

In Fig.~\ref{fig:wignerPure}, we display the Wigner quasiprobability distribution computed using the initial wavepacket in Fig.~\ref{fig:autocorr_pure_even_wavepacket}(b). The figure shows a quasiprobability flow from one centre to the other, with a strong ``quantum bridge" near $p=0$. As the time flows, there is a  motion of frequency $\omega_{20}=4.73~$a.u., which corroborates the statement that only the overlap integral between the ground and second excited state is relevant to the problem at hand. The plot corresponds to almost a whole period of the autocorrelation function, and illustrate an oscillating behavior in the Wigner quasiprobability distribution. The bridges become slanted, change slope and then return to their original configuration at $T\approx 1.33$ a.u. (not shown)\footnote{For a more thorough picture of the phase-space evolution of Wigner functions for parameters corresponding to Fig. \ref{fig:wignerPure} and \ref{fig:wignerMixed} see the following YouTube videos: \url{https://youtu.be/kA_udKIxVwM} and \url{https://youtu.be/pp1oDZ6T45k} .}

Fig.~\ref{fig:wignerMixed}, in contrast, illustrates the phase-space evolution if we use the parameters in Fig.~\ref{fig:autocorr_pure_even_wavepacket}(a). The quasiprobability flow behaves in a much more convoluted way, with additional maxima near the quantum  bridge and in both wells. For longer times, there will also be tails in the Wigner functions moving away from the potential wells, which indicate an overlap with a delocalised eigenstate, or in some cases ionisation. These tails are visible in the bottom panels of Fig.~\ref{fig:wignerMixed}. For a detailed discussion of tails of Wigner functions in the context of strong-field ionisation see our previous publications \cite{zagoya2014,chomet2019}.

From the autocorrelation function, we expect that the frequencies $\omega_{20}$ and $\omega_{40}$ will play a role.  This convoluted behavior will be discussed in the rightmost column of Fig.~\ref{fig:wignerMixed}, in which, instead of constructing Wigner quasiprobability distributions using the full analytical wavefunction, we consider only a coherent superposition
\begin{equation}
    \Psi_{20}(x,t)=\Lambda_0\exp \left(-iE_0t/\hbar\right)\psi_0(x)+\Lambda_2\exp \left(-iE_2t/\hbar \right)\psi_2(x),
    \label{eq:partialphi}
\end{equation}
between the ground and second excited state, where the overlap integrals $\Lambda_n (n=0,2)$ are given by Eq.~(\ref{eq:overlapintegrals}).  The partial Wigner quasiprobability flow mirrors the overall behavior reported in the central column of Fig.~\ref{fig:wignerMixed} except for the substructure and the tails. It determines the existence of the quantum bridges and their slopes, whose time evolution has the frequency $\omega_{20}$. This shows the dominance of this specific coupling and is expected, as tunneling should be dominated by the lower frequency. However, a modulation is introduced due to the non-vanishing overlap between the ground and the fourth excited state and its higher frequency $\omega_{40}$. Furthermore, the tails are absent in the partial results. This is due to the missing overlap integral with the fourth excited eigenstate, which is significantly broader (see Fig.~\ref{fig:eigenfunctions}).

\section{Conclusions}
\label{sec:conclusion}

In this work, we present an analytical method for solving Schr\"odinger's equation in a hyperbolic double well potential of any height and width. Our approach allows us to determine the entire eigenspectrum and corresponding eigenfunctions for the system up to an arbitrary precision, in contrast to finding only few individual eigenstates and eigenenergies in a related approach of quasi-exactly-solvable models \cite{downing2013,razavy1980exactly,agboola2014solvability}. 

By means of the exchange of variables $x \rightarrow \xi$ and $x \rightarrow \zeta$ along with exploiting the parity of the potential, Schr\"odinger's equation for the hyperbolic-double well is reduced to Heun's equation \cite{downing2013} with the resulting wavefunction involving Heun's infinite power series. Instead of truncating this series to a polynomial \cite{ronveaux,downing2013} we avoid the problem of constraining the height/width of the potential by focusing on the series' convergence. This leads to a quantisation condition which reduces a problem of finding allowed energies to finding roots of a high-degree polynomial with coefficients generated from a three-term recurrence relation. The proposed quantisation condition displays some similarities with the results of the theory of quasi-exactly-solvable models, in particular sharing the same polynomial factorisation property \cite{bender1996quasi}. However, it is more general as it gives a whole spectrum instead of a small subset of eigenvalues. 

Using the initial wavepackets with independently tunable width and peak location, we calculate the overlap integrals with the system's eigenstates in terms of incomplete gamma functions. This allows us to analytically evaluate temporal evolution of the wavepacket as a function of its initial parameters. 
This method is then employed to study tunelling through a central barrier for different initial wave packets for different coherent superpositions involving two or more eigenstates.  Apart from an excellent agreement with the numerical model in \cite{chomet2019}, which was used as a benchmark, this analytical model provides far more insight about the system's dynamics. Specifically, the autocorrelation functions and Wigner quasiprobability distributions exhibit  a periodic motion that can be precisely determined using the system's eigenfrequencies. These dynamics are strongly dependent on the width of the initial wavepacket, and the time-independent overlap integrals obtained for an eigenfunction basis has greatly facilitated our studies. In addition to that, the present phase-space studies support the conclusions in \cite{chomet2019} that the intra-center quasiprobability flows caused by quantum interference, dubbed `quantum bridges' in our previous publication, have their time evolution determined by frequencies intrinsic to the system, instead of a non-adiabatic response to an external driving field as proposed in \cite{takemoto2011}. Moreover, for the specific, field-free case studied in this article, we have determined such frequencies exactly for a hyperbolic double-well, thus going beyond the rough estimates in \cite{chomet2019}. 
                  
The analytical model developed here may form a basis for investigating a wide-range of static or time-dependent perturbative effects and be helpful in testing predictions of more realistic but non-analytically-solvable models of a double-well. In particular, the model could analytically address the issue of finding the optimal parameters for enhanced ionisation \cite{chomet2019} in a time-dependent field. For that purpose, it will be necessary to overcome a series of obstacles. First, the model developed in this article is strongly reliant on parity and inversion symmetry. Adding even a static field would break this symmetry and require changes in the way the eigenstates are calculated. Second, ionisation would require the computation of continuum states, which are not yet available in the present model. Third, a time dependent field would imply that the time-dependent Schr\"odinger equation may no longer be reduced to an eingenvalue equation. Hopefully, a low enough driving-field frequency may allow for a quasi-static picture with an effective potential and time-dependent dressed states. 

The quantisation condition proposed here may be successfully applied to a wider class of potentials than the one given by Eq. (\ref{potential_generic}). Although the arguments developed in Section \ref{recurrencerelationargument} exploit the specific form of the recurrence relation, it should be possible to extend it to the other potentials for which the Schr\"odinger equation may be solved in terms of Heun's infinite power series generated from a three-term recurrence relation. In particular we verified that for the distinct \textit{symmetric} hyperbolic potential proposed by \cite{xie2012new} the quantisation condition predicts the eigenvalues of $E=-1,-0.19113$ and $E=-1.0048, -0.25$ for parameters $\{V_1=1, V_2=-6, V_3=6\}$ and $\{V_1=1, V_2=-7, V_3=27/4\}$  respectively, in consonance with the ones earlier reported (\cite{xie2012new}, p. 4-5). Furthermore, it appears that the proposed condition may be also applied to the \textit{asymmetric} hyperbolic double-well potential \cite{hartmann2014}, predicting the eigenvalues of $E=0.311, 2.434, 3.875$ for parameters $\{w_1=15, w_2=12, w_3=1\}$ in agreement with the Wronskian's method used by \cite{hartmann2014}. However, for asymmetric wells we do not expect the wavefunctions to have even/odd parities,  Hence, a modified procedure would have to be applied to find the suitable initial wavepackets used for temporal evolution. Such asymmetric double-well potential could, for example, model the dynamical behaviour of the wavepacket in heteronuclear molecule setups. 

\section*{Acknowledgments}
We would like to thank Prof. S. Yurchenko and Dr T. Mavrogordatos for inspiring discussions and critical comments. We would like to acknowledge funding from the UK Engineering and Physical Sciences Research Council (EPSRC) (Grant EP/T019530/1).

\section*{Appendix A: Continued fraction formulation}

The continued fraction formulation of the quantisation condition presents as follows:
\begin{equation}
0=\cfrac{1}{b_0+\cfrac{a_1}{b_1+\cfrac{a_2}{ b_2+\cdots}}}
\end{equation}
where $$a_n(\alpha,\beta)=\frac{C_n(\alpha,\beta)}{A_n(\alpha,\beta)}$$ and $$b_n(\alpha,\beta)=\frac{B_n(\alpha,\beta)}{A_n(\alpha,\beta)}$$ for $n\geq1$ and $b_0=0$. Given value of $\alpha$ we can numerically search for such $\beta$ which fulfills the above condition by a process of successive approximations. The proof of the above statement is presented below.

The value of the infinite continued fraction of the form
\begin{equation*}
    \Tilde{x}=b_0+\frac{a_1}{b_1+}\frac{a_2}{b_2+} \cdots
\end{equation*} 
(following notation used by (\cite{gautschi},p. 28)) may be written as
\begin{equation*}
\Tilde{x}=\lim_{n\rightarrow \infty} \frac{M_n}{L_n},
\end{equation*}
where $$b_0+\frac{a_1}{b_1+}\frac{a_2}{b_2+} \cdots \frac{a_n}{b_n}=\frac{M_n}{L_n}.$$
In such case, $M_n$ and $L_n$ fulfill the following three-term recurrence relations (\cite{gautschi}, p. 28): 
\begin{equation*}
\begin{split}
M_n=b_n M_{n-1} + a_n M_{n-2} \\
L_n=b_n L_{n-1} + a_n L_{n-2}
\end{split}
\end{equation*}
differing only by initial conditions: $M_{-1}=1$, $M_0=0$, $L_{-1}=0$, $L_0=1$. At this point we recognise that recurrence relation for $L_n$ is equivalent to Heun's recurrence relation for $v_n$ (Eq. 2.6) if we choose $$a_n(\alpha,\beta)=\frac{C_n(\alpha,\beta)}{A_n(\alpha,\beta)}$$ and $$b_n(\alpha,\beta)=\frac{B_n(\alpha,\beta)}{A_n(\alpha,\beta)}$$ for $n\geq1$. Hence, we conclude that searching for such $\beta$ that $v_n(\alpha,\beta)=0$ for large $n$ based on (Eq. 2.6) is equivalent to searching for a root of the $1/\Tilde{x}$ infinite continued fraction (corresponding to a $n\rightarrow \infty$ limit). From there it may be easily observed that:
\begin{equation*}
0=\frac{1}{\Tilde{x}}=0+\cfrac{1}{0+\cfrac{a_1}{b_1+\cfrac{a_2}{ b_2+\cdots}}}
\end{equation*} 
which may be solved for $\beta$ by root-finding methods with level of precision set by number of terms used to approximate the infinite continued fraction.  

\printbibliography

\end{document}